\documentclass[fleqn,usenatbib]{mnras}

\usepackage{newtxtext,newtxmath}

\usepackage[T1]{fontenc}

\DeclareRobustCommand{\VAN}[3]{#2}
\let\VANthebibliography\thebibliography
\def\thebibliography{\DeclareRobustCommand{\VAN}[3]{##3}\VANthebibliography}

\usepackage{graphicx}	
\usepackage{amsmath}	
\usepackage{threeparttable}
\usepackage{subfigure}
\usepackage{makecell}
\usepackage{soul}

\title[Refining the Gaia DR3 Parallax Zero-point]{Refining the Gaia DR3 Parallax Zero-point: A Hybrid Approach Combining Global Parametric Correction with Local Refinement}

\author[Ding et al.]{
Ye Ding,$^{1,2}$
Shilong Liao,$^{1,2}$\thanks{E-mail: shilongliao@shao.ac.cn}
Zhaoxiang Qi,$^{1,2}$ \thanks{E-mail: zxqi@shao.ac.cn}
Qiqi Wu,$^{1,2}$
Qi Xu,$^{1,2}$
and Keyu Zhu$^{1,2}$
\\
$^{1}$Shanghai Astronomical Observatory, Chinese Academy of Sciences, Shanghai 200030, China\\
$^{2}$University of Chinese Academy of Sciences, Beijing 100049, China
}

\date{Accepted 2026 June 5. Received 2026 June 4; in original form 2026 May 8}

\pubyear{2026}

\begin{document}
\label{firstpage}
\pagerange{\pageref{firstpage}--\pageref{lastpage}}
\maketitle

\begin{abstract}
{The \textit{Gaia} Data Release 3 (GDR3) parallaxes are affected by a complex bias that depends on stellar magnitude, color, and celestial position, with amplitudes reaching tens of microarcseconds ($\mu$as). Standard global parametric models (e.g., Lindegren et al. 2021, hereafter L21) effectively remove large-scale trends but struggle to resolve small-scale spatial systematics due to functional rigidity.}
{We aim to construct a flexible, data-driven calibration map that eliminates these residual local systematics without imposing rigid functional forms.}
{We propose a "Global Pre-correction + Local Refinement" hybrid strategy. First, we utilize the L21 model as a baseline to remove the dominant magnitude and color-dependent biases. Second, we model the \textit{residual} zero-point using a Local Non-parametric method based on a Sliding Window technique. This approach fits local trends using $k$-nearest neighbors from quasars (for faint stars, $G \ge 18$) and wide binaries combined with Large Magellanic Cloud (LMC) (for bright stars, $G < 18$).}
{Our hybrid model demonstrates significant improvements over the standard L21 solution. Validation against different samples reveals a remarkably flat residual map with near-zero bias across the full sky. Our mathematical attempt at calibrating the parallax zero-point is expected to provide a useful reference for the zero-point correction in future \textit{Gaia} DR4, and to help move towards a physical resolution of this issue.}
\end{abstract}

\begin{keywords}
Astrometry -- Parallaxes -- Methods: data analysis -- Catalogs -- Reference systems
\end{keywords}




\section{Introduction}

The \textit{Gaia} mission has revolutionized our understanding of the Milky Way by providing astrometry for over a billion sources. Despite its high precision, the \textit{Gaia} parallaxes are affected by systematic errors, specifically the parallax zero-point offset (PZPO). \citet{2021A&A...649A...2L} presented a global offset of -17 $\mu as$, based on quasars (QSOs) in \textit{Gaia} Data Release 3 (GDR3).


Numerous studies have assessed the PZPO using diverse astrophysical tracers. In the faint and intermediate regimes, independent analyses using QSOs ($G$ $\simeq$ 14 to 21, \citealt{2021PASP..133i4501L}) , red clump stars ($G$ $\simeq$ 10 to 16, \citealt{2021ApJ...910L...5H}), eclipsing binaries ($G$ $\simeq$ 13 to 19, \citealt{2021ApJ...911L..20R}), and Galactic RR Lyrae stars ($G$ $\simeq$ 15 to 17, \citealt{2021ApJ...909..200B}) generally confirm the overall negative bias, albeit with varying mean offsets ranging from $\sim -20$ to $-30\,\mu$as.
The calibration challenge is particularly acute in the bright magnitude regime ($G \lesssim 13$), where the scarcity of absolute extragalactic references and the complexities of CCD gating schemes introduce significant uncertainties. Investigations utilizing bright tracers—such as the eclipsing binaries ($G$ $\simeq$ 5 to 12, \citealt{2021ApJ...907L..33S}) , red giant branch stars ($G$ $\simeq$ 9 to 13, \citealt{2021AJ....161..214Z}; $G$ $\simeq$ 13 to 16, \citealt{2022AJ....163..149W}) , Cepheids ($G$ $\simeq$ 6 to 11, \citealt{2021ApJ...908L...6R}), open cluster members ($G$ $\simeq$ 8 to 17, \citealt{2021A&A...649A...5F}), and binary systems with orbital parallaxes ($G$ $\approx$ 6, \citealt{2023A&A...669A...4G}; \citealt{2025AJ....169..211D} )—have yielded a wide scatter in the estimated offset. 

Crucially, these discrepancies indicate that the PZPO is not merely a global constant.
\citeauthor{2021A&A...649A...4L} (\citeyear{2021A&A...649A...4L}, hereafter L21) presented that the PZPO is a complex function of magnitude ($G$), effective wavenumber ($\nu_{\text{eff}}$), and ecliptic latitude ($\beta$). 
However, the L21 correction is sample-dependent and also shows degraded performance in specific regions, making it unsuitable for full-sky applications. \citet{2024A&A...691A..81D} investigated the PZPO within the Galactic plane — a region characterized by extreme stellar crowding and high interstellar extinction that presents unique challenges for astrometric processing. This work demonstrated that L21 correction does not apply effectively to the Galactic plane.
What's more, \citet{2021A&A...654A..20G} independently investigated the spatial dependence of the PZPO using a large sample of QSOs. By discretizing the sky with the HEALPix formalism (\citealt{2005ApJ...622..759G} )and employing local polynomial functions to parameterize the offsets, this work explicitly demonstrated that the PZPO exhibits highly complex, localized variations across the celestial sphere, confirming that simple, globally smooth functions of ecliptic latitude are insufficient to fully capture these spatial structures.

\begin{figure}
              \centering
              \subfigure{\includegraphics[width=0.8\linewidth]{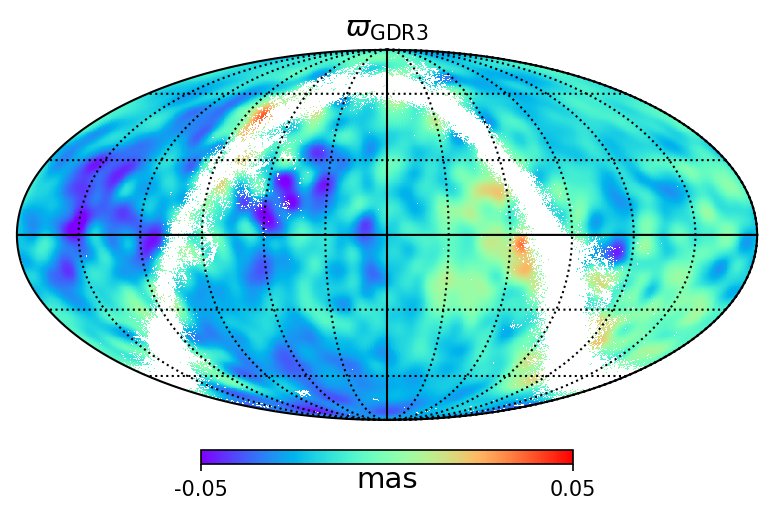}}
              \subfigure{\includegraphics[width=0.8\linewidth]{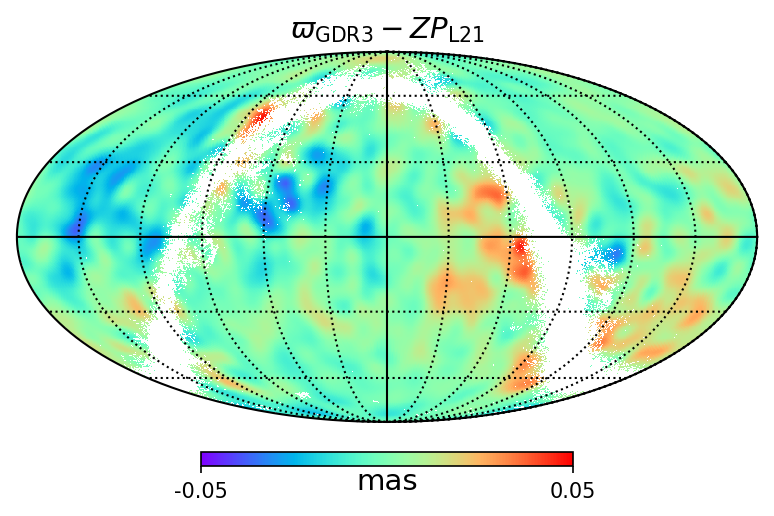}}
              \caption{All-sky distribution of the mean parallaxes for the QSOs, comparing the raw data with the L21-corrected values in Equatorial coordinates. Top panel: the GDR3 parallaxes ($\varpi_{\text{GDR3}}$). Bottom panel: parallaxes after applying the L21 correction ($\varpi_{\text{GDR3}} - ZP_{\text{L21}}$). The maps are smoothed with a Gaussian kernel of FWHM = 8 to highlight large-scale structures. }
              \label{fig:valid_l21_qso}
        \end{figure}

\begin{figure}
              \centering
              \subfigure{\includegraphics[width=0.8\linewidth]{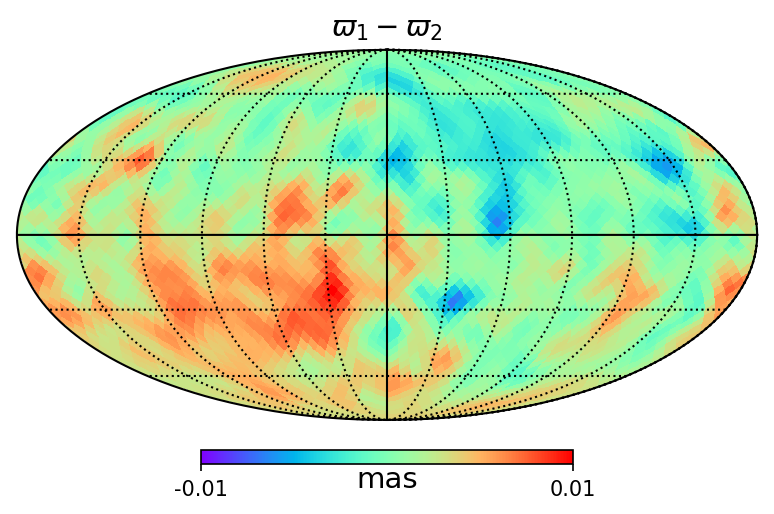}}
              \subfigure{\includegraphics[width=0.8\linewidth]{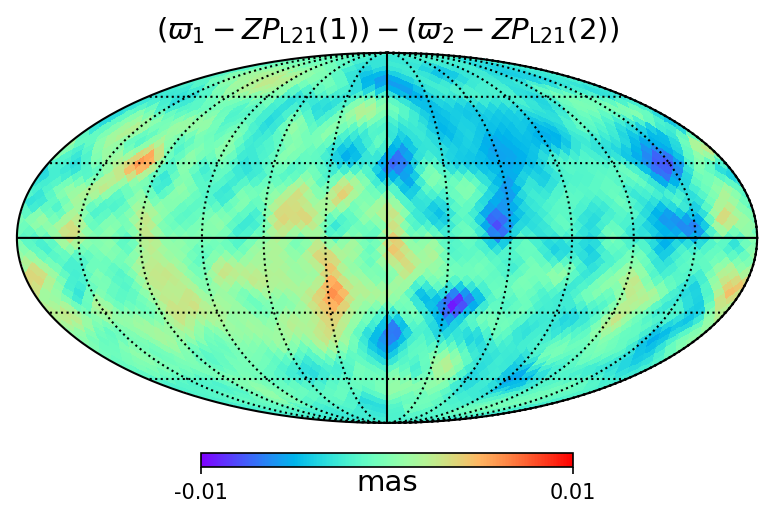}}
              \caption{All-sky distributions of mean differential parallaxes for the WBs, comparing the raw data with the L21-corrected values in Equatorial coordinates. Top panel: the GDR3 differential parallaxes between 2 components ($\varpi_{1} - \varpi_{2}$). Bottom panel: differential parallaxes after applying the L21 correction ($(\varpi_{1} - ZP_{\text{L21}}(1)) - (\varpi_{2} - ZP_{\text{L21}}(2))$). The maps are smoothed with a Gaussian kernel of FWHM = 10 to highlight large-scale structures.}
              \label{fig:valid_l21_wb}
        \end{figure}

These collective findings indicate that the PZPO can not be fully described by a global parameterized function of stellar parameters ($G, \nu_{\text{eff}}, \beta$). The conclusion is further supported by our own extensive validation of the full-sky GDR3 data. We compare the spatial distribution of parallaxes before and after applying L21 correction, using QSOs and binaries detailed in Sect. \ref{sec:data}. As illustrated in Figs.~\ref{fig:valid_l21_qso} and \ref{fig:valid_l21_wb},  prominent spatial PZPO structures stubbornly persist even after the full L21 correction is applied. Specifically, the residual maps exhibit pronounced, large-scale artifacts — with amplitudes fluctuating by $\sim 10$ to $30\,\mu$as in the significant regional patches.

In an effort to resolve the systematic, we firstly tried to improve the L21 correction using more abundant and extensive samples, including QSOs, wide binaries, and stars in the Large Magellanic Cloud (LMC). Results prove that more samples can not solve the problem completely. Second, to describe the spatial variation of the PZPO, we developed a more flexible and complicated Global Parametric Model combining B-splines and higher-order Spherical Harmonics. However, both global approaches revealed a fundamental limitation: an inability to capture spatial variations of the PZPO (see Appendix \ref{app:global_limits}).

Building upon these insights, the present work aims to develop a comprehensive, all-sky calibration framework. Since we lack access to the raw \textit{Gaia} data (e.g., CCD-level observations) required for a fundamental instrumental recalibration, this work focuses on providing a purely mathematical, empirical correction based on the published catalog data. We present a hybrid framework: "Global Pre-correction + Local Refinement." We utilize the L21 model as a baseline to remove the bulk of the parameter dependencies, and then employ a Sliding Window approach to model and remove the local PZPO residuals. Rather than attempting to force a rigid global function to capture localized features, this method strategically isolates spatial systematics for data-driven, local refinement. The resulting approach yields a seamless and highly accurate zero-point (ZP) map, unconstrained by the limitations of global parameterization.

The paper is structured as follows. In Sect.~\ref{sec:data}, we describe the selection and purification of the tracer datasets used for training and validation, including QSOs, binaries, LMC, and cluster members. Section~\ref{sec:methodology} outlines our methodological approach, detailing the construction of the global pre-correction and the local sliding window refinement. In Sect.~\ref{sec:results}, we present a comparative validation of our models, first demonstrating the limitations of purely global solutions and then establishing the superior performance of the hybrid strategy across both faint and bright magnitude regimes. Finally, we discuss the physical implications of our findings in Sect.~\ref{sec:discussion} and provide a summary and future outlook in Sect.~\ref{sec:conclusions}.

\section{Data}
\label{sec:data}

\begin{table*}
\caption{Statistics of the cross-matches of GDR3 to external QSO catalogs.}           
\label{catalogs}      
\centering  
\begin{threeparttable}
\begin{tabular}{c c c c c c }        
\hline\hline  
    {Catalog} & {Sources} & {Unique matches}  & {Non-overlapping sources} & {Filtered sources} & {Reference} \\
      \hline
    DESI DR1 & 1 772 694 & 627 580   & 214 426 & 30 287 & \cite{2026AJ....171..285D} \\  
    SDSS DR16Q & 750 414 & 467 549  & 38 532 & 1241 &\cite{2020ApJS..250....8L} \\
    LAMOST DR10 & 83 376 & 69 728  & 3218 & 394 &\cite{2023ApJS..265...25J} \\
    LQAC-6 & 1 758 557 & 1 738 455 & 39 236 & 324 &{\cite{2024A&A...683A.112S}} \\
    Quaia &1 295 502 & 1 295 497  & 151 236 & 116 261 &{\cite{2024ApJ...964...69S}} \\
    GPQ & 162 488 & 102 133 & 20 418 & 8148 & \cite{2021ApJS..254....6F,2022ApJS..261...32F, 2025ApJS..278....6H}\\ 
    CatGlobe & 1 889 813 & 1 889 813  & 170 121 & 83 266 & \cite{2024ApJS..271...54F, 2025ApJS..279...54F} \\
    KiDS DR4 & 1 095 711 & 74 965 & 11 768 & 862 & {\cite{2021A&A...649A..81N}} \\
    Gaia-CRF3 & 1 614 173 & 1 614 173  & 1 614 173 & 1 215 941 & {\cite{2022A&A...667A.148G}} \\[2ex]
    ALL & -- & -- & 2 263 128 & 1 456 562 & -- \\
                         
    \hline                                  
    \end{tabular}
\begin{tablenotes}
\footnotesize
    \item Notes. The external catalogues are listed in Col. 1. Column 2 shows the total number of sources in each catalogue. Column 3 provides the number of unique matches in GDR3 for each catalogue. Column 4 presents the number of matched sources satisfying Eqs. (\ref{eq1}). Column 5 shows the number of sources retained after filtering. Column 6 provides references for the catalogs. 
\end{tablenotes} 
        
\end{threeparttable}
    
\end{table*}

To investigate the systematic, we require extensive and high-quality samples, including QSOs, binaries, stars in the LMC, and globular clusters. We utilized 4 distinct datasets for model training and validation:

\begin{enumerate}
    \item QSOs: Serving as the absolute reference for the faint regime. Table \ref{catalogs} lists the result of cross-matches with 8 external catalogs followed by further astrometric filtering. First, we cross-matched GDR3 astrometric catalog with all external catalogues and then retained the closer sources where more than one \textit{Gaia} source is matched to a source in an external catalog. Second, we flagged the sources in external catalogs which are overlapped with Gaia-CRF3 (\citealt{2022A&A...667A.148G},  hereafter GCRF3). Third, we combined all catalogs in the order of GCRF3, DESI DR1, SDSS DR16Q, LAMOST DR10, LQAC-6, Quaia, GPQ, CatNorth, and KiDS DR4, in the process of removing overlapped sources between these catalogs. The final union contains 2,263,128 sources,  and the numbers of the retained sources from different catalogs are given in Column 4. 
    
    Strict filtering was applied to remove outliers, which are adopted from the GCRF3 quasar selection \citep{2022A&A...667A.148G}. 
    
    \begin{align}\label{eq1}
        \left\{     
                \begin{aligned}
                (i) \quad &astrometric\_params\_solved = 31, \\
                (ii) \quad &\left | (\varpi +0.017 mas )\right |/ {\sigma }_{\varpi }<5,  \\
                (iii) \quad &{{\mathrm{X}}_{\mu}}^{2} \equiv  \begin{bmatrix}\mu_{\alpha*} & \mu_{\delta} \end{bmatrix}{Cov(\mu )}^{-1}\begin{bmatrix} \mu_{\alpha*} \\ \mu_{\delta} \end{bmatrix} < 25.
                \end{aligned}
        \right.
        \end{align}
        
    Criterion (i) restricts the sample to sources with full five-parameter astrometric solutions; Criterion (ii) imposes a statistical significance cut on the parallax. We reject sources whose observed parallax significantly deviates from the global PZPO of GDR3 ($\sim -17\,\mu$as); and Criterion (iii) evaluates the proper motion kinematics. We compute the error-normalized proper motion magnitude, $\mathrm{X}_{\mu}^2$, which rigorously accounts for the uncertainties ($\sigma_{\mu\alpha*}$, $\sigma_{\mu\delta}$) and their correlation via the inverse covariance matrix, $\mathbf{Cov}(\mu)^{-1}$. By demanding $\mathrm{X}_{\mu}^2 < 25$, we robustly filter out foreground stellar contaminants that exhibit statistically significant transverse motion.
    
    1,456,562 sources survived the filtering. The numbers of Non-overlapping sources that survived the filtering by Eq. (\ref{eq1}) are provided in column 5 of Table \ref{catalogs}. 
    \\

    \item Physical Binaries: Physical binaries offer valuable insights into determining the PZPO, as the binary components have essentially near-identical parallaxes. Used to transfer the PZPO to the bright regime, we selected high-probability wide binaries (WBs) from \cite{2021MNRAS.506.2269E}. The catalog of WBs contains 1.8 million candidate physical binaries. The sample is restricted to the subset of 1 256 400 pairs with >90\% probability of being bound (chance alignment <10\%). and further quality filtering is described below. 
    
    \begin{align}\label{eq2}
        \left\{     
                \begin{aligned}
                (i)\quad &astrometric\_params\_solved1,2 = 31 , \\
                (ii)\quad &\left|(\varpi_{1}-\varpi_{2}) \right| / \sqrt{{{\sigma }_{\varpi_{1}}}^{2}+{{\sigma }_{\varpi_{2}}}^{2}} < 5, \\
                (iii)\quad & {\varpi_{1,2}} < 0.5 \quad mas, 
                \end{aligned}
        \right.
    \end{align}
    where `1' and `2' denote the components with the brighter and the fainter G magnitude, respectively. Criterion (i) requires both components to possess full five-parameter astrometric solutions; Criterion (ii) enforces a statistical consistency check on the differential parallax. Assuming the true parallax difference of a physical pair is negligible, we reject pairs where the observed difference $(\varpi_1 - \varpi_2)$ deviates by more than $5\sigma$ from zero, effectively eliminating severe astrometric outliers or optical (chance-alignment) pairs; Criterion (iii) imposes a strict data quality threshold, retaining only components with formal parallax uncertainties $\sigma_\varpi < 0.5$ mas to maintain a high signal-to-noise ratio in the differential measurements. 483, 639 physical pairs survived the filtering. About 60\% pairs were removed, mostly due to Criterion (i).
    \\

    \item LMC and globular clusters: 
    Resolved stellar systems, such as the LMC and Galactic globular clusters, provide invaluable datasets for calibrating the offset in the bright and extreme color regime. Their utility stems from a key assumption: all member stars within a given system share the same true (though a priori unknown) mean parallax. Consequently, spatial variations or parameter-dependent trends in the observed parallaxes of members can be directly attributed to systematic errors.

    The distance modulus of the LMC, $(m-M)_{0} =18.49 \pm 0.09$ mag \citep{2014AJ....147..122D}, corresponds to a parallax of 20.04 $\pm$ 0.83 $\mu as$. The LMC field is rich in stellar density but heavily contaminated by foreground Milky Way stars. To isolate a clean sample of LMC members, we applied a rigorous multi-stage filtering process based on astrometric quality indicators and kinematic properties, as detailed in Appendix~\ref{app:lmc_selection}. This selection yielded a core sample of 1,145,280 high-probability members. 
    

    We utilized 10 most populous Galactic globular clusters from the catalog of \citet{2021MNRAS.505.5978V}, as detailed in Table \ref{tab:clusters}. 
    It's important to note that the parallaxes of the member stars that contribute to the mean parallax of a cluster have been corrected by L21 recipe. We enforced strict selection criteria , retaining only stars with full five-parameter solutions, high membership probability ($>90\%$), and the high astrometric reliability flag ($qflag=3$). These clusters provide the datasets for the validation. 

\end{enumerate}

\section{Methodology}
\label{sec:methodology}
Our approach to calibrate the PZPO was driven by an iterative process of model refinement and diagnosis.
First, to mitigate the sample dependence of the L21 correction, we attempted to use more expanded dataset covering the full-sky, including QSOs especially the Galactic plane, WBs, and LMC stars, aiming to leverage the rich information in the datasets. Second, to overcome the inability to describe the PZPO spatial structures of the L21 correction, we developed a more complicated global parametric model, combining B-splines and higher-order Spherical Harmonics. 


\subsection{Refinement of the Standard L21 Model}
\label{sec:refinement_l21}
Our initial strategy was to investigate whether the sample dependence of the L21 correction could be mitigated by utilizing more extensive and diverse training datasets. The L21 calibration primarily relied on three distinct classes of tracers to map the PZPO variations: QSOs from Gaia-CRF3 to anchor the faint end, sources in the LMC to constrain color dependencies, and physical binaries to bridge the calibration to the bright regime ($G$ < 13). 
To improve the parameter constraints, particularly in challenging regions, we expanded the QSO sample by incorporating approximately 240,000 sources beyond the official GCRF3 catalog (summarized in Table~\ref{catalogs}). Crucially, this included the GPQ catalog, which substantially improves the full-sky coverage by filling the critical gaps along the Galactic plane. Furthermore, we incorporated a large, independent sample of WBs. Unlike the binary sample in L21, which was predominantly used for the brightest stars, our WB sample provides robust differential constraints across a broader magnitude range, specifically bolstering the intermediate regime ($13 < G < 16$). 
By integrating this significantly enhanced collection of QSOs, WBs, and our newly constructed LMC star sample, we retrained the PZPO correction using the exact functional forms provided by L21.

The L21 model parametrizes the ZP bias $ZP(G, \nu_{\text{eff}}, \beta)$ as a linear combination of basis functions:
\begin{align}\label{eq:bias_function}
        ZP(G, \nu_{\text{eff}},\beta) = \displaystyle\sum_{j} \displaystyle\sum_{k} q_{jk}(G) \, c_{j}(\nu_{\text{eff}}) \, b_{k}(\beta),
\end{align}
where the functions 
\begin{align}\label{eq:bias_function}
        q_{jk}(G) = \displaystyle\sum_{i} z_{ijk} \, g_{i}(G), \; j = 0...4, \; k = 0...2
\end{align}
are piecewise linear in G. Among the functions, $g_{i}(G)$ are the basis functions in magnitude, $c_{j}(\nu_{\text{eff}})$ are the basis functions in color, and $b_{k}(\beta)$ are the basis functions in ecliptic latitude. The coefficients $z_{ijk}$ are the free parameters used to fit $Z$ to the given data. For more details on the basis functions, see L21 Appendix A.

We adopted the exact basis function definitions and knot positions ($G=6.0, 10.8, \dots, 21.0$) as specified in the L21 implementation to ensure structural compatibility. Crucially, rather than adopting the step-by-step sequential fitting procedure used in the L21 calibration, we determined the L21 model coefficients through a simultaneous joint least-squares optimization across our entire combined dataset. 
For WBs,  we directly model the differential offset ($\Delta ZP$). While L21 first anchored the parallax bias of the fainter component using QSO and LMC data before deriving the bias for the bright primary, our approach utilizes the observed differential residuals as the direct target for localized fitting. 
In the treatment of the LMC sample, we initially adopted the methodology of L21, treating the mean parallax of the LMC ($\varpi_{\text{LMC}}$) as an unknown free parameter to be fitted alongside the ZP coefficients.

However, in the context of our simultaneous joint fit, we discovered that leaving $\varpi_{\text{LMC}}$ unconstrained resulted in significant unphysical drift. The optimization process yielded a mean parallax that deviated substantially from the expected geometric distance ($\varpi_{\text{LMC}} \approx 20.04\,\mu$as). This indicates a strong degeneracy within the joint likelihood landscape: the unconstrained LMC parallax was unduly influenced by the relative constraints imposed by the WBs and QSO samples, effectively absorbing broader systematic trends rather than reflecting the true distance. To break this degeneracy and ensure a robust calibration, we abandoned the free-parameter approach and fixed the mean parallax of the LMC sample to $\varpi_{\text{LMC}} = 20.04\,\mu$as.

Retraining the global parametric model with our expanded and diverse dataset yielded measurable improvements in characterizing specific parameter-dependent biases. As illustrated in Figs.~\ref{fig:global_comparisons_qso}, \ref{fig:global_comparisons_binary}, and \ref{fig:global_comparisons_lmc}, the application of our retrained model (green circles) substantially mitigates the systematic offsets in the parallax residuals for QSOs, WBs, and LMC sources when compared to the raw, uncorrected data (black circles), bringing the residuals considerably closer to the true zero-point across the $G$ and $\nu_{\text{eff}}$ domains. However, a critical limitation emerges when evaluating the model's performance in the spatial domain. As shown in Figs.~\ref{fig:global_comparisons_maps_refinedl21}, despite the optimized coefficients, significant large-scale spatial systematics stubbornly persist across the celestial sphere. This outcome proves that the rigid functional form adopted by the L21 model is fundamentally insufficient to capture the complex spatial variations of the PZPO.

\subsection{Investigation of a Global Parametric Model}
\label{sec:global_limits}
To address the limitations of the L21 formulation, we developed a more complicated global parametric model to capture the complex spatial variations of the PZPO. This model employed an additive combination of cubic B-splines (to model $G$ and $\nu_{\text{eff}}$ dependencies with higher flexibility) and higher-order Spherical Harmonics (SH, up to degree $l=10$) to better capture spatial variations. Spherical harmonics offer a flexible and orthogonal basis for decomposing spatial signals on the celestial sphere. Their multiscale nature allows low-order terms to capture broad, large-scale trends (e.g., dipole, quadrupole), while higher-order terms can effectively model localized, small-scale structures. Compared to pixel-based or other global parametric functions, spherical harmonics avoid overfitting to noise when properly truncated and provide a natural framework for separating instrumental signatures from astrophysical signals. In the context of Gaia astrometric calibration, as discussed in \cite{2012A&A...547A..59M} and \cite{2021PASP..133b4501L,2021PASP..133i4501L}, spherical harmonics have been successfully used to capture both global and local spatial systematics, thereby improving the homogeneity of the reference frame across the entire sky. The detailed mathematical formulation and the process of fitting is provided in Appendix~\ref{app:global_model}. We constrained this model simultaneously using our full suite of tracers: QSOs, WBs, and LMC stars. Similar to Sect. \ref{sec:refinement_l21}, we fitted the differential offset ($\Delta ZP$) directly for WBs and fixed the mean parallax of the LMC sample to $\varpi_{\text{LMC}} = 20.04\,\mu$as.

While this higher-order B-spline and Spherical Harmonic (BSH) model successfully removed the global mean offsets across various magnitude and color regimes—performing comparably well to the refined L21 model—it fundamentally failed to resolve the complex spatial systematics. As illustrated by the red circles in Figs.~\ref{fig:global_comparisons_qso}, \ref{fig:global_comparisons_binary}, and \ref{fig:global_comparisons_lmc}, the BSH model achieves similar success to the refined L21 approach in mitigating parameter-dependent biases, bringing the residuals considerably closer to the true zero-point across the $G$ and $\\nu_{\text{eff}}$ domains. However, this increased parametric flexibility does not translate to improved spatial resolution. As shown in Figs.~\ref{fig:global_comparisons_maps_bsh}, significant residual structures stubbornly persist across the all-sky maps.
This outcome highlights a critical and fundamental limitation: globally smooth functions are inherently ill-suited to model the localized features of the scanning law. Simply increasing the complexity of a global parametric model (e.g., by adding more SH terms) yields diminishing returns. It introduces the risk of overfitting the noise in well-sampled regions without effectively capturing the underlying complex spatial structures. This crucial realization firmly motivated our shift away from purely global fitting paradigms toward a flexible, data-driven, localized refinement strategy.

\subsection{The Hybrid Strategy: Global Pre-correction + Local Refinement}
\label{sec:hybrid_strategy}

Given the inherent limitations of global parametric functions in modeling the spatial systematics, we adopted a third and final strategy: a purely mathematical, data-driven correction framework. Our final approach is a hybrid strategy that decouples the problem into two distinct stages: a global pre-correction to handle large-scale parameter dependencies, followed by a local refinement to address the remaining spatial residuals.

Based on the comparative analysis of global parametric models, we determined that while global functions fundamentally fail to resolve complex spatial structures, they remain essential for modeling the large-scale, complex dependencies on magnitude and color. 

To establish the optimal baseline for our hybrid framework, we selected our Refined L21 model over the original L21 calibration and the higher-order BSH model. The rationale is twofold: First, the specific basis function architecture of L21 (piecewise linear in $G$ coupled with color polynomials) provides a more stable and less prone-to-overfitting parameterization compared to the dense Spherical Harmonics expansions of the BSH approach. Second, and more importantly, by retraining the L21 coefficients on our comprehensive dataset—crucially incorporating the dense color-magnitude constraints from expanded QSO, WB,and the LMC sample—we effectively mitigated the sample-dependency biases inherent in the official release. As demonstrated in Figs.~\ref{fig:global_comparisons_qso}-\ref{fig:global_comparisons_lmc}, the Refined L21 model offers a superior, flattened parameter-space baseline prior to spatial refinement.

Therefore, our final Hybrid Strategy proceeds in two decoupled stages:

\subsubsection{Step 1: Global Pre-correction}
We utilized the standard L21 model as a baseline pre-processor. For every source in our training sets, we calculate and subtract the L21 predicted PZPO. This step removes the dominant trends associated with magnitude and color, "flattening" the ZP landscape. The target for our subsequent local modeling is the \textit{residual parallax}, $\delta \varpi$:

\begin{equation}
    \label{eq:res}
     \delta \varpi = \varpi_{\text{GDR3}} - \varpi_{\text{true}} - ZP_{\text{L21}}^{\text{Ref}}(G, \nu_{\text{eff}}, \beta)
\end{equation}
where $\varpi_{\text{GDR3}}$ is the observed parallax, $\varpi_{\text{true}}$ represents the true absolute parallax of the tracer, and $ZP_{\text{L21}}^{\text{Ref}}$ is our Refined L21 correction model. For the extragalactic QSO sample, $\varpi_{\text{true}}$ is strictly defined as zero; for Galactic tracers such as binaries, we rely on their differential properties.

\subsubsection{Step 2: Local Refinement via Sliding Window}
\label{sec:local_refinement}

To model the spatially correlated residuals without introducing artificial boundaries inherent to fixed-grid methods, we employ a continuous, non-parametric Sliding Window (SW) technique. For any target coordinate on the sky, we define a highly localized empirical model driven exclusively by its nearest neighbors in the training set. 

Within each local window, we approximate the residual ZP ($\delta \varpi$) as a low-order polynomial function of the stellar parameters: magnitude ($G$) and effective wavenumber ($\nu_{\text{eff}}$). To ensure absolute numerical stability during the least-squares optimization, we perform a rigorous standardization of the local variables. The standardized variables, $\tilde{G}$ and $\tilde{\nu}$, are defined as:
\begin{equation}
    \tilde{G} = \frac{G - \langle G \rangle}{\sigma_G}, \quad \tilde{\nu} = \frac{\nu - \langle \nu \rangle}{\sigma_\nu}
    \label{eq:standardization}
\end{equation}
where the weighted mean ($\langle \cdot \rangle$) and weighted standard deviation ($\sigma$) are calculated dynamically from the specific subset of neighbors within the current window. The local ZP residual is then modeled as:
\begin{equation}
    ZP_{\text{res}}(\tilde{G}, \tilde{\nu}) = \sum_{i=0}^{D} \sum_{j=0}^{D-i} p_{ij} \tilde{G}^i \tilde{\nu}^j
    \label{eq:local_poly}
\end{equation}
where $D$ is the polynomial degree, and $p_{ij}$ are the unknown coefficients optimized via a robust, weighted least-squares fit. 

A critical challenge is the drastic variation in calibrator density across the sky. To guarantee robustness, our SW algorithm employs an adaptive strategy. For a given target, it initially searches for $N_{\text{min}}$ neighbors within a fixed radius. If insufficient neighbors are found, it switches to a $k$-Nearest Neighbors ($k$-NN) search, capped by a maximum tolerated radius ($R_{\text{max}}$) to prevent severe non-local biases. Furthermore, the polynomial degree $D$ is selected dynamically: while dense regions may support a linear plane ($D=1$), sparse or ill-conditioned regions automatically downgrade to a simpler, conservative weighted mean ($D=0, ZP_{\text{res}}=p_{00}$).  Crucially, to ensure numerical stability against localized outliers and preserve the spatial fidelity of the correction, the weighted least-squares fit is augmented with two mechanisms: (1) a Gaussian distance kernel that assigns higher statistical weight to physically closer neighbors, and (2) an iterative $3\sigma$-clipping procedure that actively identifies and rejects highly discrepant residual measurements during the optimization loop.

We apply this robust framework to construct two distinct, continuous residual maps based on different tracer populations.

\begin{enumerate}
    \item The QSO-Anchored Map.
    This map provides the fundamental absolute reference for our calibration, as it is anchored by a purified sample of $\sim 1.4 \times 10^6$ extragalactic QSOs ($\varpi_{true} \equiv 0$). For any given target coordinate, the algorithm initially searches for a minimum of $N_{\text{min}}=50$ neighbors within a fixed physical radius of $R=5^\circ$. If the local density is insufficient, the search expands up to a hard upper limit of $R_{\text{max}}=10^\circ$. Given the generally dense and homogeneous sky coverage of the QSO sample, the local data subset typically supports the robust fitting of a multi-dimensional linear plane ($D=1$, optimizing the coefficients $p_{00}, p_{10}, p_{01}$). 
    \\

    \item The Bright-Tracer Map.
    To provide robust spatial constraints in the bright magnitude regime ($G < 18$), where the availability of absolute QSO calibrators drops precipitously, we construct a secondary map using a composite sample of Galactic tracers. This sample integrates differential constraints from WBs.
    For the dataset of $\sim 4.8 \times 10^5$ WB pairs, we adopt the approximation that the observed differential residual is overwhelmingly dominated by the bright primary component (i.e., assuming the systematic residual of the faint secondary is negligible after the global L21 pre-correction). Consequently, the target calibration signal for the primary star is defined directly by the differential residual: $ZP_{\text{res}}^{\text{Bright}}(1) \approx \delta\varpi_1 - \delta\varpi_2$. This approach yields a dense, full-sky network of relative spatial constraints.
    
    Before modeling, all derived bright calibrators are combined and subjected to rigorous purification. We retain only sources satisfying $G < 18$, possessing derived absolute residuals $|ZP_{\text{res}}| < 500\,\mu$as, and having propagated measurement errors $\sigma < 50\,\mu$as. Utilizing this purified catalog, the SW algorithm constructs the final continuous calibration field by fitting a local linear polynomial ($D=1$, optimizing the coefficients $p_{00}, p_{10}, p_{01}$) with a neighborhood requirement of $N_{\text{min}}=30$, $R=5^\circ$, and a maximum search radius $R_{\text{max}}=15^\circ$. This configuration effectively produces a spatially smoothed, yet structure-preserving, zero-point map tailored for bright stars.
    
    
    
\end{enumerate}

\subsubsection{Final Application}

The final ZP correction for any given source combines the parameter-dependent baseline from the global model with the localized spatial refinement from our sliding window approach. The final corrected parallax, $\varpi_{\text{corr}}$, is expressed as:
\begin{equation}
    \varpi_{\text{corr}} = \varpi_{\text{GDR3}} - \left[ ZP_{\text{L21}}^{\text{Ref}}(G, \nu_{\text{eff}}, \beta) + ZP_{\text{res}}(\alpha, \delta, G, \nu_{\text{eff}}) \right]
    \label{eq:final_correction}
\end{equation}

To evaluate the local residual term $ZP_{\text{res}}$, we must judiciously select between our two trained maps. While the QSO-Anchored map spans fainter magnitudes, its predictions for $G$ <17 are perilous extrapolations due to data scarcity. Conversely, the Bright-Tracer map is specifically engineered for $G < 18.0$.

Crucially, the evaluation of the local residual term, $ZP_{\text{res}}$, is strictly partitioned according to the source's magnitude to ensure robust, data-driven predictions and avoid unphysical extrapolations. 
We define a strict physical hard-boundary at $G = 18.0$.
It is important to clarify that this $G=18.0$ threshold is distinct from the instrumental transition points at $G \approx 13$ and $16$. While the latter dictate the underlying physical error patterns, our $G=18.0$ boundary is fundamentally a data-driven algorithmic limit. It demarcates the regime where the spatial density of the ultimate absolute reference—extragalactic QSOs—becomes too sparse to support robust local non-parametric fitting. 

To maximize the calibration accuracy for diverse astrophysical studies, we recommend a targeted application strategy based on the nature of the sources being analyzed. If the target sources are known to be QSO, we highly recommend evaluating the residual term $ZP_{\text{res}}$ exclusively using the QSO-Anchored Map, regardless of their apparent magnitude. Although bright QSOs ($G < 18$) are sparse, their astrometric behavior is intrinsically identical to the training set of this map, ensuring the most physically consistent absolute ZP correction. For all other Galactic stellar sources (e.g., Cepheids, RR Lyrae, field stars), we suggest the transition from the QSO-anchored map to the Composite Bright-Tracer map.
\begin{itemize}
    \item Faint Sources ($G \ge 18.0$): The residual term $ZP_{\text{res}}$ is evaluated exclusively using the QSO-Anchored Map. \footnote{In the Galactic plane, the extreme scarcity of QSOs limits the spatial resolution of this map. Users focusing specifically on faint sources at very low Galactic latitudes are advised to apply the specialized corrections derived in \citet{2024A&A...691A..81D}.}
    \item Bright Sources ($G < 18.0$): The residual term $ZP_{\text{res}}$ is evaluated exclusively using the Bright-Tracer Map.
\end{itemize}

      \begin{figure}
              \centering
              \subfigure{\includegraphics[width=0.9\linewidth]{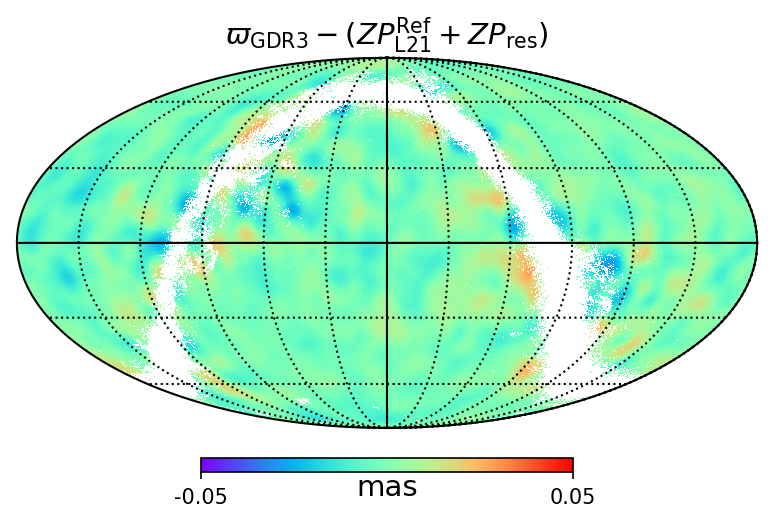}}
              \subfigure{\includegraphics[width=0.9\linewidth]{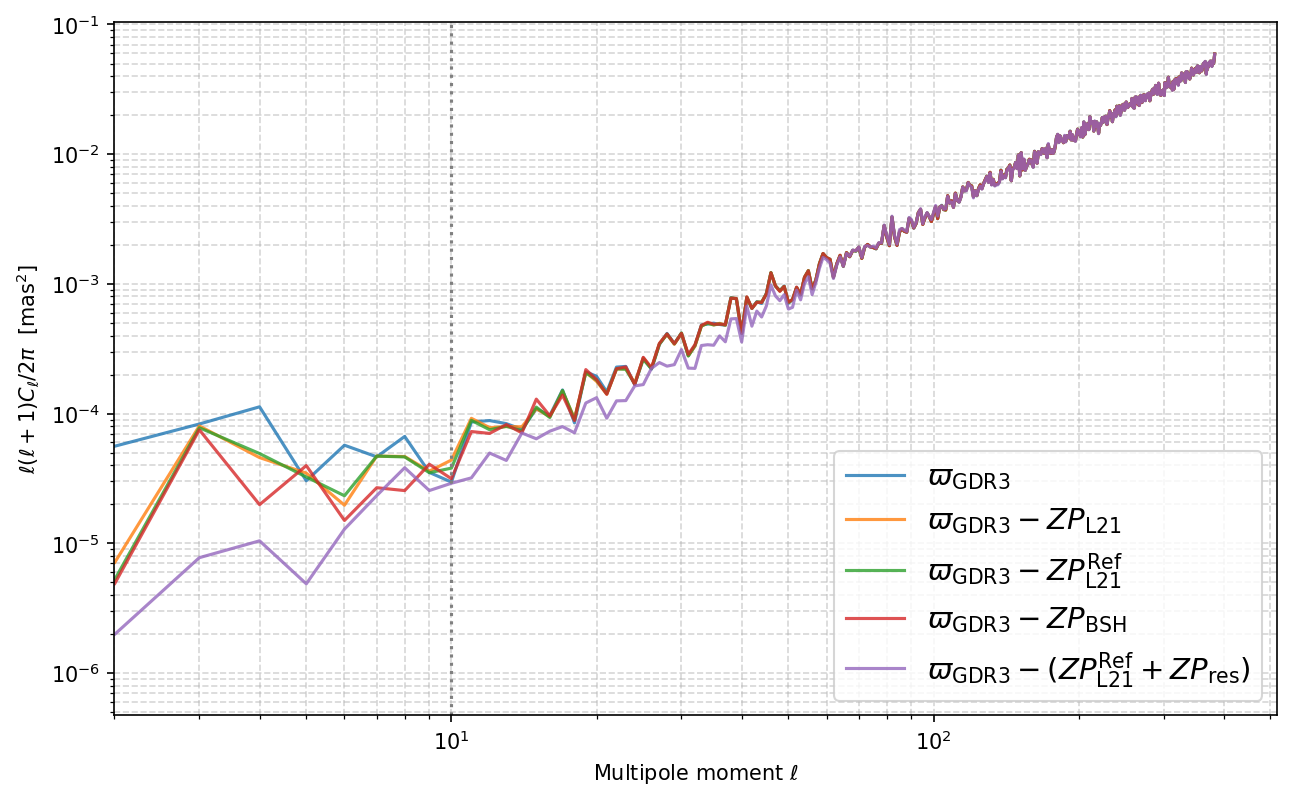}}
              \caption{Validation of the Hybrid Model in the Faint Regime using QSOs. Top: All-sky distribution (in Equatorial coordinates) of the parallax residuals after applying our final hybrid correction ($\varpi_{\text{GDR3}} - (ZP_{\text{L21}}^{\text{Ref}} + ZP_{\text{res}})$). The map is smoothed with a $8^\circ$ Gaussian kernel. The resulting field is highly uniform and centered on zero, demonstrating the visual removal of scanning-law artifacts. 
              Bottom: The angular power spectrum (APS) of the residuals. The vertical axis displays the power $\ell(\ell+1)C_\ell/2\pi$ in mas$^2$. While global parametric models (orange, green, red lines) fail to suppress spatially correlated errors at large-to-intermediate scales, the hybrid model (purple line) reduces the residual power by nearly an order of magnitude across this entire broad frequency range, confirming the successful removal of scanning-law artifacts and broad regional biases without overfitting the small-scale noise ($\ell > 100$). }
              \label{fig:hybrid_maps_qso}
        \end{figure}
        
        \begin{figure}
              \centering
              \subfigure{\includegraphics[width=0.9\linewidth]{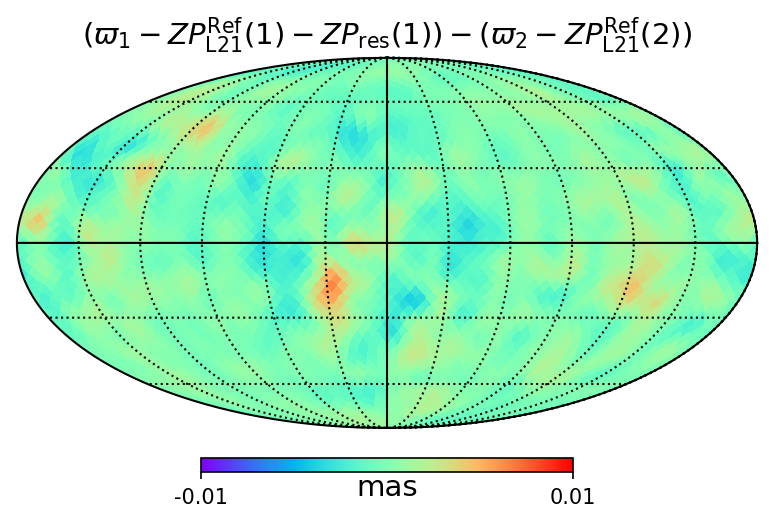}}
              \subfigure{\includegraphics[width=0.9\linewidth]{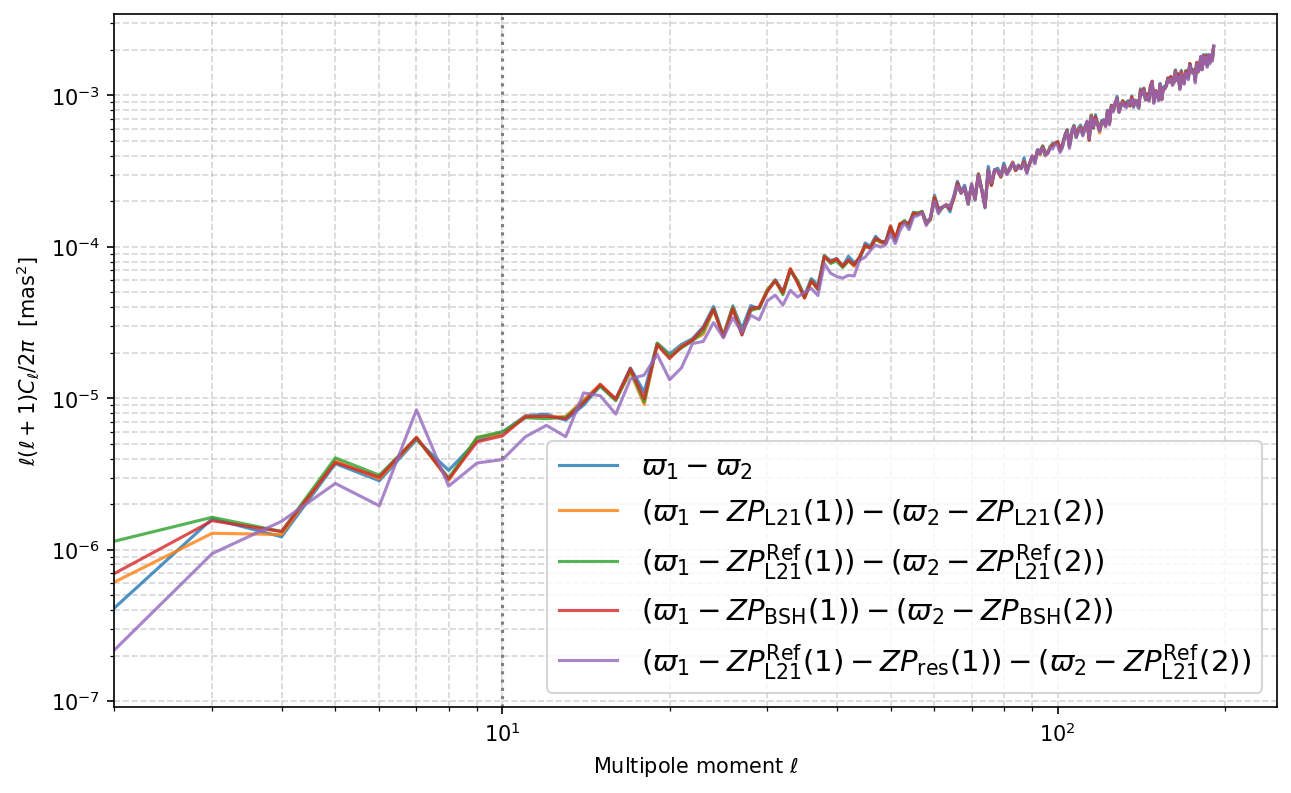}}
              \caption{ Validation of the Hybrid Model in the Faint Regime using the WBs. Top: All-sky distribution (in Equatorial coordinates) of differential parallax residuals after applying the hybrid correction independently to each binary component ($ (\varpi_{1} - ZP_{\text{L21}}^{\text{Ref}}(1) - ZP_{\text{res}}(1) ) - (\varpi_{2} - ZP_{\text{L21}}^{\text{Ref}}(2))$). The map is smoothed with a $10^\circ$ Gaussian kernel. 
              Bottom: The APS of the differential residuals. Despite the independent application of the local correction, the hybrid model (purple line) achieves an lower power spectrum than the global models (L21, Refined L21, BSH) across large and intermediate spatial scales. This demonstrates that the local refinement map for bright stars successfully captures true spatial systematics without inflating decorrelation noise, underscoring the robustness of the sliding-window filtering strategy.}
              \label{fig:hybrid_maps_wb}
        \end{figure}

        \begin{figure*}
              \centering
              \subfigure{\includegraphics[width=0.33\linewidth]{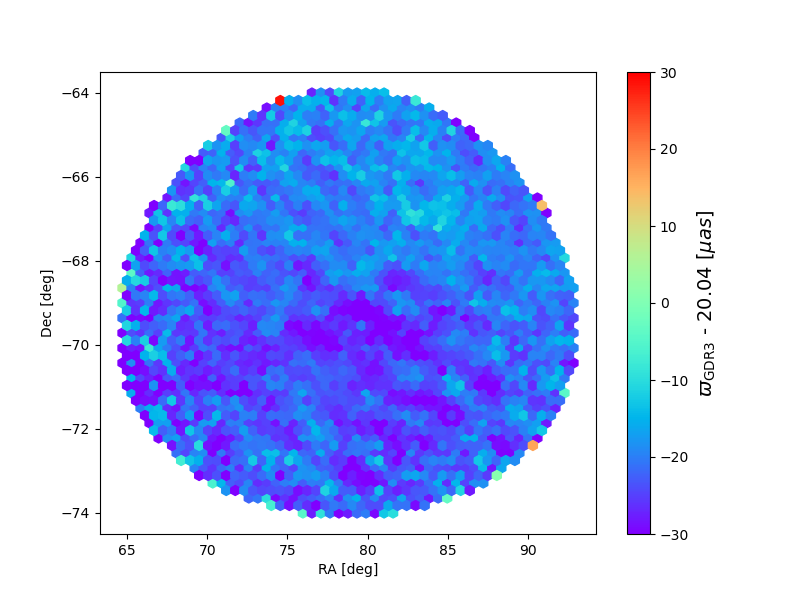}}
              \subfigure{\includegraphics[width=0.33\linewidth]{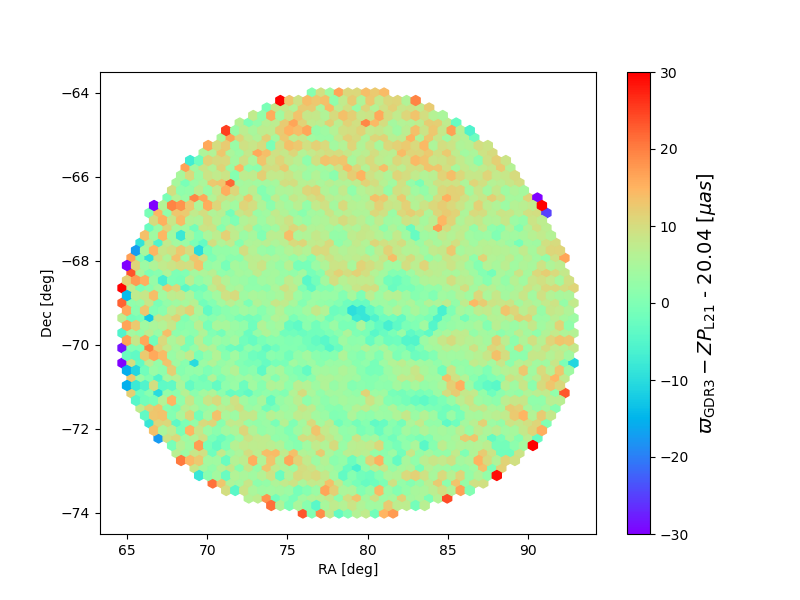}}
              \subfigure{\includegraphics[width=0.33\linewidth]{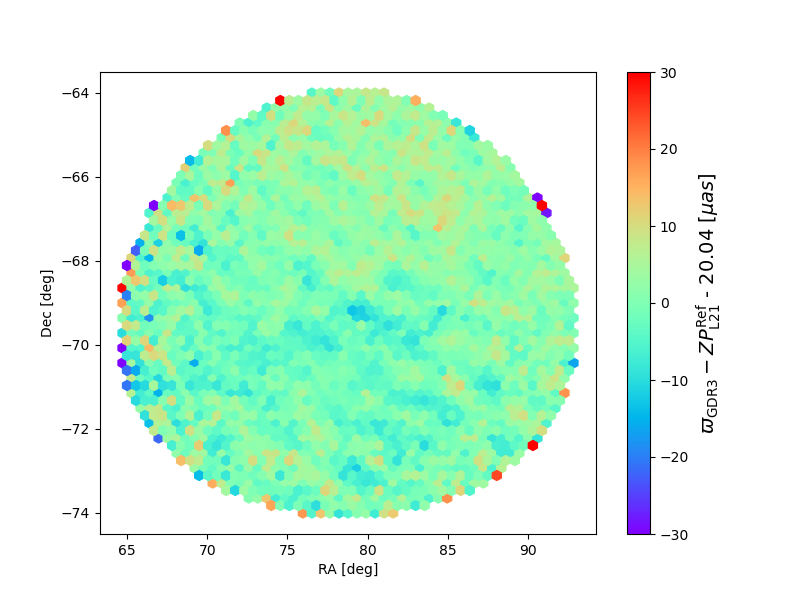}}
              \caption{ Spatial distribution of parallax residuals for the LMC under three scenarios: Raw (left), L21 correction (center), and our Hybrid correction (right). }
              \label{fig:lmc}
        \end{figure*}

\section{Results and Validation}
\label{sec:results}

In this section, we evaluate the efficacy of our proposed calibration strategies. We begin by diagnosing the performance of global parametric models. The details are described in Appendix ~\ref{app:global_limits}, highlighting the persistent systematic errors that motivated our new approach. We then present the results of our final Hybrid Model in this Section, demonstrating its ability to resolve these systematics. 

To rigorously evaluate the efficacy of our "Global Pre-correction + Local Refinement" strategy across the celestial sphere, we computed the angular power spectrum (APS, \citealt{1973ApJ...185..413P}; \citealt{2003ApJS..148..135H}) of the residual maps for both faint (QSOs) and bright (WBs) validation samples.  The APS provides a highly sensitive metric for spatially correlated errors; an ideal, bias-free dataset containing only random measurement noise would exhibit a flat "white noise" spectrum at high $\ell$, with minimal power at low $\ell$ modes.

The top panel of Fig.~\ref{fig:hybrid_maps_qso} displays the all-sky distribution of the hybrid-corrected residuals, $\varpi_{\text{GDR3}} - (ZP_{\text{L21}}^{\text{Ref}} + ZP_{\text{res}})$. In stark contrast to the maps produced by the L21 correction (as shown previously in the bottom panel of Fig.~\ref{fig:valid_l21_qso}), the hybrid correction yields a remarkably flat and uniform zero-point landscape. The prominent large-scale regional biases and the "striping" artifacts that could be associated with the Gaia scanning law (e.g., \citealt{2021AJ....161...58F, 2021A&A...649A...5F}) have been visually eradicated, with the residuals fluctuating tightly around zero across the entire celestial sphere. 

This visual improvement is quantitatively confirmed by the APS in the bottom panel. The raw GDR3 parallaxes (blue line) exhibit substantial excess power across all large and intermediate angular scales ($\ell \lesssim 40$), indicative of prominent spatial systematics. While the global parametric models (L21, Refined L21, and BSH) offer some reduction in power, they consistently fail to fully resolve these broad regional biases and leave significant residual structures at intermediate scales ($10 \lesssim \ell \lesssim 40$) potentially associated with the scanning law.
In stark contrast, our final hybrid correction (purple line) achieves a dramatic and comprehensive suppression of residual power across all these critical spatial scales. By incorporating the data-driven local refinement, the spatial variance is reduced by nearly an order of magnitude compared to the best global models, effectively flattening the power spectrum down to the fundamental white-noise floor for $\ell \lesssim 40$. Furthermore, the convergence of all curves at very small angular scales ($\ell > 100$) confirms that our local refinement successfully eliminates systematic structures while properly preserving the intrinsic, uncorrelated measurement noise of individual sources without overfitting.

To evaluate the performance of our hybrid model in the bright magnitude regime, where absolute astrometric anchors like quasars are scarce, we turn to the differential parallaxes of WBs. For a true physical pair, the intrinsic parallax difference should be zero; thus, the observed differential parallax, $\Delta\varpi_{\text{obs}} = \varpi_1 - \varpi_2$, directly probes the differential PZPO. 
Crucially, to rigorously validate our calibration framework, the application of the hybrid model to the binary set must perfectly mirror the assumptions made during the construction of the Bright Star Residual Map. Specifically, we maintain the approximation that the observed differential residual is overwhelmingly dominated by the bright primary component. Therefore, the local spatial refinement ($ZP_{\text{res}}$) is applied exclusively to the primary star, while both components receive their respective global parameter-dependent baseline corrections ($ZP_{\text{L21}}^{\text{Ref}}$). The final hybrid-corrected differential residual is thus formulated as:
\begin{equation}
    \Delta\varpi_{\text{resid}}^{\text{Hybrid}} = \left( \varpi_1 - ZP_{\text{L21}}^{\text{Ref}}(1) - ZP_{\text{res}}(1) \right) - \left( \varpi_2 - ZP_{\text{L21}}^{\text{Ref}}(2) \right)
\end{equation}
This independent, asymmetric application represents an exceptionally stringent test of the local model's spatial stability, as any spurious noise or overfitting in the $ZP_{\text{res}}(1)$ map would severely degrade the differential precision and amplify the residual variance.

The top panel of Fig.~\ref{fig:hybrid_maps_wb} shows the all-sky distribution of this final differential residual. In marked contrast to the highly structured L21-corrected map (Fig.~\ref{fig:valid_l21_wb}, bottom), the hybrid-corrected map appears substantially smoother and more homogeneous. The severe, large-scale artifacts have been effectively mitigated, indicating that the local residual map successfully captured and subtracted the true spatial systematics of the bright primary without introducing unphysical decorrelation noise.
This visual assessment is quantitatively corroborated by the APS in the bottom panel. Remarkably, despite the asymmetric application of the local correction, the hybrid model (purple line) achieves a consistently lower power spectrum compared to all purely global models (L21, Refined L21, BSH) across large and intermediate angular scales. The fact that the addition of a local residual term—derived from a relatively sparse and noisy network of bright calibrators—actually decreases the spatial correlation of the differential residuals compared to highly-smoothed global functions is a profound testament to the efficacy of our robust sliding-window filtering strategy. It confirms that the bright-star local refinement map reliably captures genuine spatial systematics without artificially inflating small-scale noise.

To rigorously test the absolute accuracy of our model in the bright regime, which binary differential tests cannot fully reveal—we turn to external validators: LMC members and a set of Galactic globular clusters.

We first examine the spatial distribution of parallax residuals for the LMC, presented in Figure~\ref{fig:lmc}. The raw parallax residuals (left panel) exhibits a severe, uncorrected negative bias ($\sim -25.44\,\mu$as) across the entire field. The application of the official L21 correction (center panel) substantially improves this distribution; however, a visually discernible, systematic positive bias remains, resulting in a residual weighted mean offset of $+3.1\,\mu$as. 
In contrast, the right panel of Fig.~\ref{fig:lmc} demonstrates the performance of our final hybrid strategy in this region. As detailed in Sect.~\ref{sec:hybrid_strategy}, because the LMC data was explicitly utilized as a dense, absolute geometric anchor during the global pre-correction stage (training the Refined L21 baseline), the resulting spatial distribution is remarkably uniform and centered tightly around zero. The weighted mean residual for the LMC is reduced to an unprecedented $-0.19\,\mu$as. This strong alignment suggests that our strategy of incorporating high-fidelity localized samples into the global parametric fit successfully eradicates the regional absolute PZPO that persist in the standard L21 model. Furthermore, because this region is effectively "flattened" by the pre-correction, the application of any subsequent local residual model (derived from sparse, external wide binaries) is intentionally suppressed, avoiding the introduction of unphysical decorrelation noise into this densely sampled, pristine region.

Beyond the LMC, we evaluated the calibration performance across a broader parameter space using the Galactic globular clusters. Table~\ref{tab:clusters} provides a comprehensive summary of the weighted mean parallax residuals for these 10 clusters across four different calibration scenarios. It is important to note that the literature mean parallaxes ($\varpi_{\text{true}}$) provided by \citet{2021MNRAS.505.5978V} were derived after applying the standard L21 zero-point corrections to the member stars. Consequently, the L21-corrected residuals in Table~\ref{tab:clusters} are intrinsically close to zero by definition. Overall, our Hybrid Correction demonstrates highly robust performance, maintaining the systematic bias at near-zero levels comparable to the L21 baseline for the vast majority of the clusters. This confirms that our local spatial refinement does not introduce spurious global offsets.

\begin{figure*}
              \centering
              \subfigure{\includegraphics[width=0.9\linewidth]{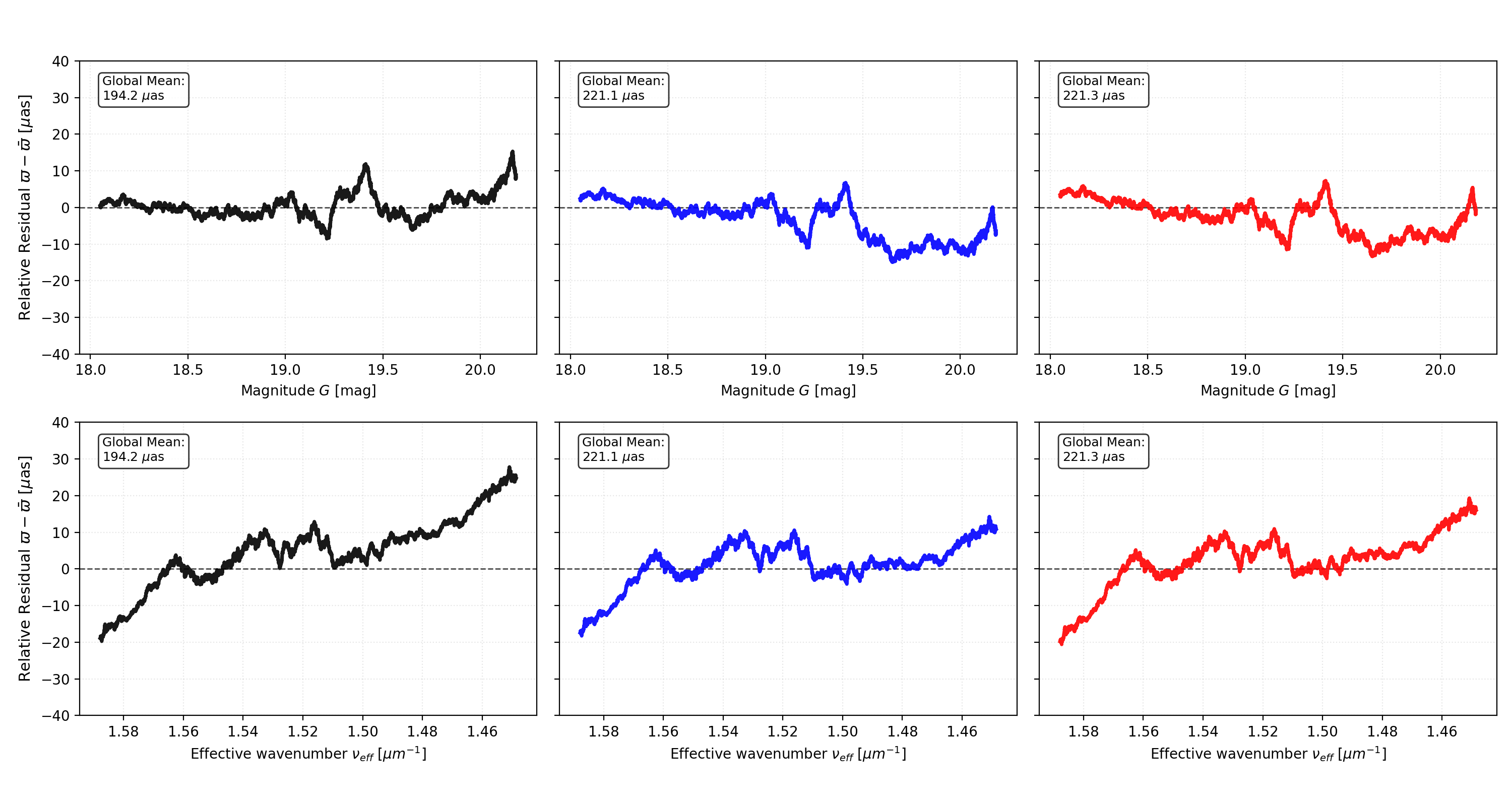}}
              \caption{ Parameter-dependent parallax residuals for the globular cluster NGC 104 with G > 18 (N=27, 519). The mean-subtracted relative residuals are plotted as a function of magnitude $G$ (top row) and effective wavenumber $\nu_{eff}$ (bottom row). Left: Raw GDR3 data exhibiting strong parameter dependencies. Center: Residuals after the L21 global correction. Right: Residuals after applying our final Hybrid model (Refined L21 + Local Sliding Window).}
              \label{fig:ngc104_trends}
\end{figure*}

However, examining the overall mean offset is insufficient to evaluate the model's ability to flatten parameter-dependent biases within a single cluster. To investigate this, we analyzed the mean-subtracted relative residuals as a function of magnitude ($G$) and color ($\nu_{\text{eff}}$). 
Figure~\ref{fig:ngc104_trends} illustrates this diagnostic for the NGC 104. As expected, the raw data (left) exhibits strong magnitude and color dependencies. While both the L21 model (center) and our Hybrid model (right) alter the overall magnitude of these trends, neither successfully flattens the curves. Strikingly, the parameter-dependent residual profiles produced by our local Hybrid model are morphologically identical to those of the L21 model, inheriting the same systematic undulations across the $G$ and $\nu_{\text{eff}}$ domains. This crucial observation indicates a fundamental limitation in applying generalized zero-point models to globular clusters. This limitation may stem from several factors: a more complex color–magnitude dependence of the \textit{Gaia} PZPO in these dense systems, unresolved correlations among the astrometric parameters of member stars due to extreme crowding, and uncertainties in cluster membership.

\begin{table*}
\caption{Weighted mean parallax residuals ($\mu$as) for globular clusters.}
\label{tab:clusters}
\centering
\begin{threeparttable}
\begin{tabular}{l c c c c c c}
\hline\hline

Sample & sources & \makecell{Mean (uncorrected parallax \\
offset) ($\mu as$)}  & \makecell{L21 correction \\
($\mu as$)} & \makecell{Refined L21 correction \\ ($\mu as$)} & \makecell{BSH correction \\ ($\mu as$)} & \makecell{Hybrid correction \\ ($\mu as$)}\\
\hline
NGC 5139  & 52 854 & -38.30 & -2.64 & -0.78 & -14.08 & -1.64 \\
NGC 104 & 39 761 &  -36.69 & -4.50 & -6.47 & +1.48 & -0.69 \\
NGC 6397 & 12 308 &  -32.10 & +3.34 & +5.62 & +1.65 & +6.29 \\
NGC 3201 & 11 391 &  -38.44 & -5.37 & -5.52 & -13.20 & -12.19 \\
NGC 6752 & 16 155 &  -34.11 & +0.12 & +0.88 & -1.44 & +2.63 \\
NGC 5272 & 8 941 &  -30.31 & +0.53 & +5.68 & -6.30 & +2.70 \\
NGC 5904 & 9 141 &  -29.77 & +2.58 & +7.73 & +0.46 & +2.81 \\
NGC 6205 & 9 537 &  -23.22 &  +2.14 & +6.12 & +3.67 & +3.87 \\
NGC 6254 & 7 130 &  -37.08 & -2.01 & +3.53 & -3.45 & -4.90 \\
NGC 6809 & 7 420 & -34.64 & +1.52 & +6.08 & -4.51 & +4.92 \\
\hline
\end{tabular}
\begin{tablenotes}
\footnotesize
    \item Notes. This table presents a selected subset comprising the 10 most populous Galactic globular clusters from \citet{2021MNRAS.505.5978V}, chosen to ensure high statistical robustness in the mean residual calculations. Column 1 lists the cluster identifier. Column 2 gives the number of member sources surviving the quality filtering. Column 3 shows the raw, uncorrected weighted mean PZPO relative to the literature distances. Columns 4-7 display the residual bias remaining after the application of four distinct calibration models: the official L21 model, our Refined L21 model, our BSH model, and our final Hybrid correction strategy, respectively.
\end{tablenotes} 
\end{threeparttable}
\end{table*}

\section{Discussion}
\label{sec:discussion}

The results presented in Sect.~\ref{sec:results} demonstrate that moving from rigid global parametrizations to a data-driven local refinement strategy significantly improves the calibration of \textit{Gaia} parallaxes. Here we discuss the implications of these findings, the strengths of the hybrid approach, and the remaining challenges in the bright regime.

A key insight from this work is the necessity of decoupling global parameter dependencies from local spatial variations. Global models like L21 are highly effective at removing the large-scale trends with magnitude and color, which are likely driven by basic instrumental properties (e.g., chromaticity, charge transfer inefficiency, basic angle variations). However, they fundamentally fail to capture the complicated small scale spatial structures (as seen in Figs.~\ref{fig:valid_l21_qso} and \ref{fig:valid_l21_wb}).
Our hybrid strategy succeeds because it respects this hierarchy: L21 acts as a "whitening" filter, flattening the parameter space, while the Sliding Window acts as a spatial high-pass filter, targeting the residual geometric systematics. The close‑to‑flat residual map of the quasar (the top panel of Fig.~\ref{fig:hybrid_maps_qso}) validates this two-step approach for the faint regime.

As highlighted by \citet{2021A&A...649A...4L}, instrumental transitions within the Gaia detectors—specifically the activation of CCD gating and changes in window class configurations at $G \approx 11, 13,$ and $16$—introduce distinct, magnitude-dependent discontinuities in the PZPO. In our hybrid framework, the mitigation of these specific, non-linear parametric jumps is entirely relegated to the Global Pre-correction stage. By retaining the piecewise linear basis functions in magnitude with strategically placed knots (including nodes specifically positioned near $13.1$ and $15.9$), our Refined L21 model is explicitly designed to absorb these sharp, magnitude-dependent steps prior to any spatial smoothing. The efficacy of this approach is evident in the top panel of Fig.~\ref{fig:global_comparisons_binary}, which demonstrates that the Refined L21 correction (green line) successfully neutralizes the abrupt ZP jumps at $G \approx 13$ and $16$ present in the raw data. By stripping away these complex magnitude dependencies first, we attempt that only the pure, small-scale spatial artifacts remain for the subsequent sliding-window refinement to resolve. 

However, a revealing insight from our validation using globular clusters is the persistent parameter-dependent structure observed in the residuals. As demonstrated in Fig.~\ref{fig:ngc104_trends}, while our hybrid strategy successfully suppresses the spatial noise across the general celestial sphere (e.g., in the QSO sample), it does not flatten the internal magnitude and color dependencies within compact clusters like NGC 104. Instead, the hybrid model produces residual trends that are essentially identical to the globally L21 baseline.

This behavior highlights a critical limitation inherent to any generalized, all-sky zero-point calibration model. Global models—including the official L21, our Refined L21, and the underlying pre-correction of our hybrid model—are trained on a vast, diverse population of field stars (QSOs, binaries, etc.). The resulting coefficients reflect an average mapping of the instrumental biases across the broad parameter space populated by these field sources.
Globular clusters, however, are highly homogeneous populations with distinct and tightly constrained distributions in the color-magnitude diagram (e.g., prominent Red Giant Branches and Horizontal Branches). The specific sequence of $(G, \nu_{\text{eff}})$ combinations present in a given cluster is often poorly represented by the global average. Consequently, when a generalized parameterization is applied to the highly specific stellar sequence of a cluster, minor imperfections in the global fit manifest as distinct, uncorrected systematic trends. 

Furthermore, our local spatial refinement ($ZP_{\text{res}}$), which operates primarily via localized spatial smoothing, is mathematically incapable of decoupling and removing these inherent parameter-dependent artifacts once they are passed down from the global pre-correction stage.

Our findings strongly suggest that relying on generalized, all-sky calibration maps is insufficient for high-precision astrometric studies of individual dense clusters. To achieve truly flat residuals across the cluster member sequence, it is imperative to develop cluster-specific zero-point models. Such models would not rely on external field stars but would be self-calibrated internally—for example, by enforcing the condition that all high-probability members, regardless of their position on the color-magnitude diagram, must yield the same absolute distance. Until such bespoke calibrations are routinely implemented, one must remain cautious of residual parameter-dependent biases when interpreting the astrometry of cluster members using generalized zero-point corrections.

\section{Conclusions}
\label{sec:conclusions}

In this work, we presented a novel hybrid framework for calibrating the Gaia EDR3 parallax zero-point. We moved beyond rigid global parametric fits by implementing a "Global Pre-correction + Local Refinement" strategy, utilizing a Sliding Window algorithm to capture local systematics.

The combination of L21 (for removing global color/magnitude trends) and a local non-parametric model (for removing spatial residuals) is a powerful approach. It effectively decouples the complex dependencies of the zero-point bias.
For faint stars ($G \ge 18$), our model achieves close alignment with the absolute quasar reference frame, removing patterns that persist in the standard L21 solution.
For bright stars ($G < 18$), while the model successfully reduces the bias in the LMC to $\sim -0.05\,\mu$as, it shows variable performance in Galactic globular clusters. This highlights the sensitivity of the binary-based transfer method to tracer density and the underlying assumptions regarding companion residuals.

While this work pushes the limits of empirical calibration based on published catalog data, the ultimate solution to the parallax zero-point problem lies beyond purely mathematical post-processing.
With the anticipated release of \textit{Gaia} DR4, which will include epoch astrometry and time-series data, the community will gain access to a deeper level of information. Future efforts should shift focus from empirical "patching" of the final catalog to a more physics-based calibration. Access to raw telemetry and CCD-level observations will allow for a direct modeling of the instrumental effects—such as basic angle variations \citep{2017A&A...603A..45B, 2018IAUS..330..231L}—that may be among the physical drivers of these systematics. By linking the astrometric residuals directly to the instrument state rather than just sky position or magnitude, we can hope to achieve a truly bias-free astrometric frame, unlocking the full potential of Gaia for high-precision astrophysics.

\section*{Acknowledgements}
We appreciate the insightful advice provided by the referee, Prof. Floor van Leeuwen. This work was supported by the National Key R$\&$D Program of China (Grant No.2025YFA1614104), the National Natural Science Foundation of China (NSFC) through grants 12173069, the Strategic Priority Research Program of the Chinese Academy of Sciences, Grant No.XDA0350205, the Youth Innovation Promotion Association CAS with Certificate Number 2022259, the Talent Plan of Shanghai Branch, Chinese Academy of Sciences with No.CASSHB-QNPD-2023-016, the International Partnership Program of the Chinese Academy of Sciences with Grant No.018GJHZ2025032FN. We acknowledge the science research grants from the China Manned Space Project with NO. CMS-CSST-2021-A12 and NO.CMS-CSST-2021-B10. This work has made use of data from the European Space Agency (ESA) mission \textit{Gaia} (\href{https://www.cosmos.esa.int/gaia}{https://www.cosmos.esa.int/gaia}), processed by the \textit{Gaia} Data Processing and Analysis Consortium (DPAC, \href{https://www.cosmos.esa.int/web/gaia/dpac/consortium}{https://www.cosmos.esa.int/web/gaia/dpac/consortium}). Funding for the DPAC has been provided by national institutions, in particular the institutions participating in the \textit{Gaia} Multilateral Agreement. This research has made use of the VizieR catalogue access tool and the cross-match service provided by CDS, Strasbourg.

This research used data obtained with the Dark Energy Spectroscopic Instrument (DESI). DESI construction and operations is managed by the Lawrence Berkeley National Laboratory. This material is based upon work supported by the U.S. Department of Energy, Office of Science, Office of High-Energy Physics, under Contract No. DE–AC02–05CH11231, and by the National Energy Research Scientific Computing Center, a DOE Office of Science User Facility under the same contract. Additional support for DESI was provided by the U.S. National Science Foundation (NSF), Division of Astronomical Sciences under Contract No. AST-0950945 to the NSF’s National Optical-Infrared Astronomy Research Laboratory; the Science and Technology Facilities Council of the United Kingdom; the Gordon and Betty Moore Foundation; the Heising-Simons Foundation; the French Alternative Energies and Atomic Energy Commission (CEA); the National Council of Humanities, Science and Technology of Mexico (CONAHCYT); the Ministry of Science and Innovation of Spain (MICINN), and by the DESI Member Institutions: www.desi.lbl.gov/collaborating-institutions. The DESI collaboration is honored to be permitted to conduct scientific research on I’oligam Du’ag (Kitt Peak), a mountain with particular significance to the Tohono O’odham Nation. Any opinions, findings, and conclusions or recommendations expressed in this material are those of the author(s) and do not necessarily reflect the views of the U.S. National Science Foundation, the U.S. Department of Energy, or any of the listed funding agencies.

\section*{Data Availability}

To facilitate the application of the PZPO corrections developed in this work, we provide the complete set of trained model coefficients, the localized spatial residual maps, and a Python package, \texttt{gaiadr3\_pzpo}. This package offers a streamlined interface to apply both the Refined L21 global pre-correction and the subsequent local Sliding Window refinement, automatically handling the magnitude-dependent application logic described in Sect.~\ref{sec:hybrid_strategy}.

The \texttt{gaiadr3\_pzpo} package, along with the necessary data files and usage tutorials, is available and hosted on GitHub at \url{https://github.com/yedings/gaiadr3_pzpo.git}. 
For long-term archival stability and reproducibility, the exact version of the data products corresponding to this article has been deposited on Zenodo under the DOI: \href{https://doi.org/10.5281/zenodo.20039676}{10.5281/zenodo.20039676}.




\bibliographystyle{mnras}
\bibliography{myref} 




\appendix

\section{Construction of the LMC sample}
\label{app:lmc_selection}
The Large Magellanic Cloud (LMC) plays a dual role in this work. First, its dense stellar population provides a rich dataset for validating the absolute ZP accuracy in a localized region. Second, we leverage the internal properties of LMC stars to construct a unique training set of differential pairs, which provides critical constraints on the color- and magnitude-dependent variations of the zero-point at a fixed sky position.

\begin{figure*}
   \centering
   \subfigure{\includegraphics[width=0.44\linewidth]{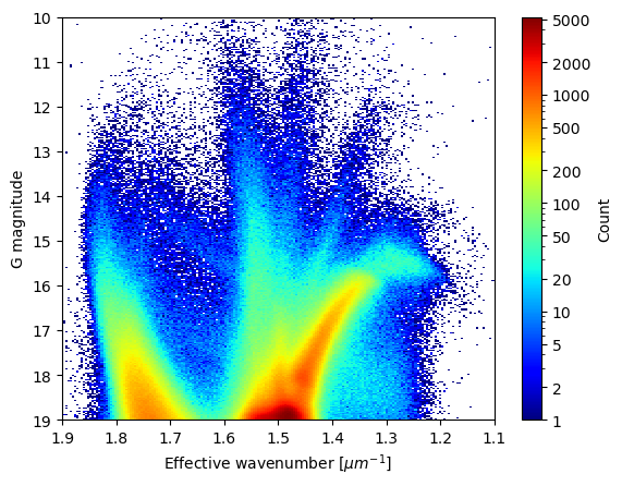}} 
   \subfigure{\includegraphics[width=0.44\linewidth]{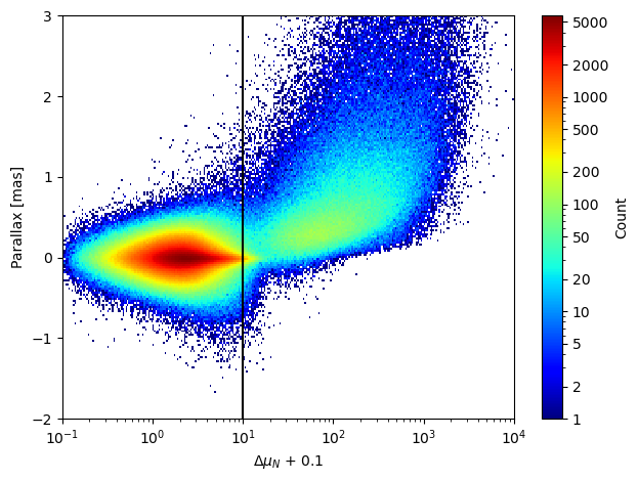}}
   \subfigure{\includegraphics[width=0.44\linewidth]{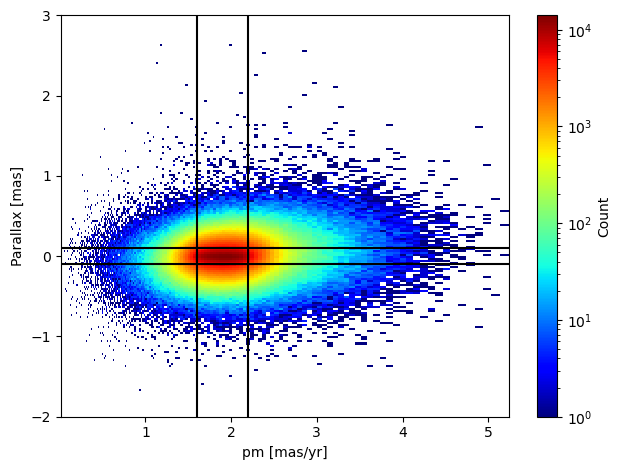}}
   \subfigure{\includegraphics[width=0.44\linewidth]{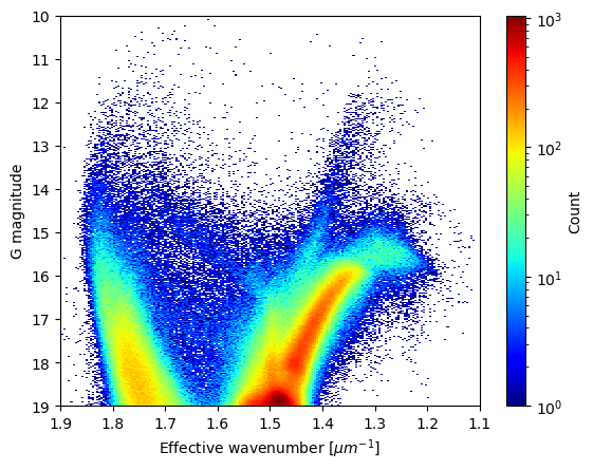}}
   \caption{Construction of the LMC validation sample. \textbf{Top left:} CMD of the initial sample. \textbf{Top right:} Distribution of parallax vs. normalized proper motion difference $\Delta\mu_N$. The vertical line indicates the cut at $\Delta\mu_N < 10$. \textbf{Bottom left:} Distribution of proper motion vs. parallax. \textbf{Bottom right:} The clean CMD of the  final refined sample ($N \approx 1.15 \times 10^6$).}
   \label{fig:lmc_selection}
\end{figure*}

We started with a cone search centered on the LMC dynamical center ($\alpha, \delta = 78.77^\circ, -69.01^\circ$) with a radius of $5^\circ$. We selected sources with five-parameter solutions, $G$ < 19, and $ruwe$ < 1.4. This initial selection yielded a sample of approximately 2.7 million sources. The color-magnitude diagram (CMD) of this sample (Fig.~\ref{fig:lmc_selection}, top left) clearly shows the characteristic features of the LMC population, but also significant foreground contamination.

To separate LMC members from foreground stars, we utilized proper motions. Following the approach of \citet{2021A&A...649A...4L}, we calculated the normalized proper motion difference, $\Delta\mu_N$, for each source (see Eq.(B.2) in Appendix B). The top right panel of Fig.~\ref{fig:lmc_selection} shows the distribution of the parallax versus $\Delta\mu_N$ of these sources. We applied a cut of $\Delta\mu_N < 10$ to remove stars with proper motions significantly deviating from the LMC mean. This step reduced the sample to approximately 2.5 million sources. The distribution of the proper motion versus parallax of these sources is shown in the bottom left panel of Fig.~\ref{fig:lmc_selection}.

To further purify the sample for high-precision zero-point validation, we applied stricter cuts on proper motion and parallax. As shown in the bottom left panel of Fig.~\ref{fig:lmc_selection}, we only selected the sources distributed in the core area. We restricted the total proper motion to the range $1.6 < \mu < 2.2$ mas yr$^{-1}$ and the observed parallaxes to the range $-0.1 < \varpi < 0.1$ mas, tightly bracketing the LMC's kinematic core. Given the LMC's true parallax of $\sim 0.019$ mas, this cut effectively removes foreground dwarfs (large positive parallax) and background artifacts while retaining the LMC members and their associated measurement scatter. The final sample consists of 1,145,280 sources. As shown in the bottom right panel of Fig.~\ref{fig:lmc_selection}, this sample exhibits a clean CMD distribution, making it an ideal dataset for mapping the spatial variations of the ZP.

\section{Details of the Global Parametric Model (BSpline SH)}
\label{app:global_model}
Our Global Parametric Model aims to provide a unified analytical description of PZPO across the entire sky and parameter space. We model the PZPO as an additive combination of basis functions that capture the dependencies on stellar parameters (magnitude $G$, effective wavenumber $\nu_{\text{eff}}$) and celestial position (ecliptic longitude $\lambda$, ecliptic latitude $\beta$). In contrast to the L21 model, which includes multiplicative cross‑terms among color, magnitude, and position, our model adopts an additive form without such interactions. We tested selected cross terms and found that they did not improve the correction significantly for our data, while increasing the risk of overfitting. Hence, we opted for the simpler representation.
The ZP function is defined as:
\begin{equation}
\label{eq:bspline_sh}
\begin{split}
    ZP(G, \nu_{eff}, \lambda, \beta) = \sum_{i=1}^{M} c_{i} B_i(G) + \sum_{j=1}^{N} c_{j} B_j(\nu_{eff}) \\
    + \sum_{l=0}^{L_{max}} \sum_{m=-l}^{l} c_{lm} Y_{lm}(\lambda, \beta)
\end{split}
\end{equation}
where  $B_i(G)$ are cubic B-spline basis functions defined on the magnitude vector $G$. We utilize a knot vector tailored to the data density, specifically concentrating knots at critical magnitude transitions (e.g., $G \approx 13$ and $16$). For this work, we used the knot sequence: [6.0, 10.8, 11.2, 11.8, 12.2, 12.9, 13.1, 15.9, 16.1, 17.5, 19.0, 20.0, 21.0]. $B_j(\nu_{\text{eff}})$ are cubic B-spline basis functions for the effective wavenumber, with knots placed at $[1.3, 1.4, 1.5, 1.7]\,\mu\text{m}^{-1}$. $Y_{lm}(\lambda, \beta)$ are the real Spherical Harmonics of degree $l$ and order $m$. We tested expansions up to $L_{max} = 10$, corresponding to $(L_{max}+1)^2 = 121$ spatial coefficients. $c_{i}$, $c_{j}$, and $c_{lm}$ are the unknown coefficients to be determined.

This additive form assumes that the dependencies on magnitude, color, and position are separable. While this reduces the number of free parameters compared to a full tensor-product model (like L21), it provides a flexible global baseline.
To intuitively illustrate the flexibility and constraints of our global parametric formulation, we visualize the three sets of basis functions defined in Eq.~\ref{eq:bspline_sh}. 
\begin{figure}
              \centering
              \subfigure{\includegraphics[width=1.0\linewidth]{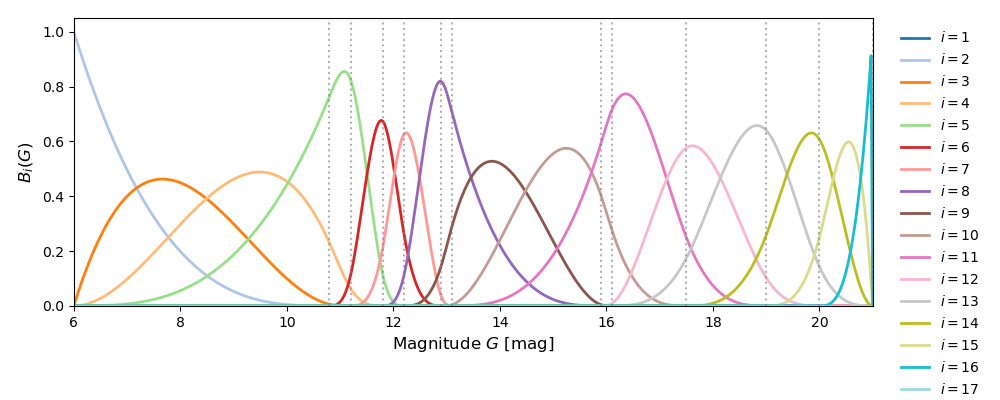}}
              \subfigure{\includegraphics[width=1.0\linewidth]{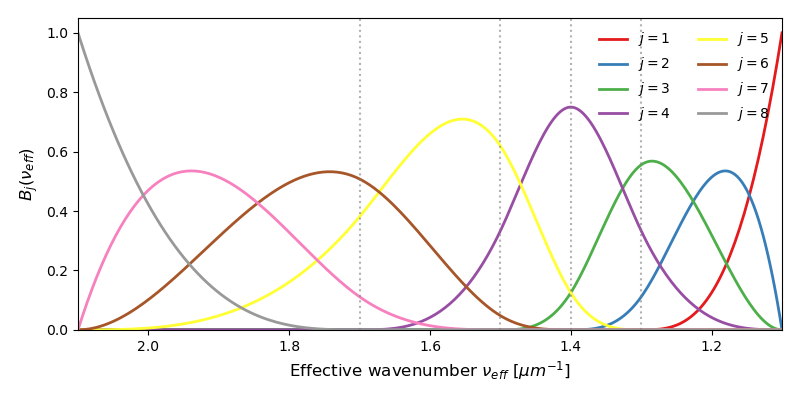}}
              \caption{
               One-dimensional basis functions for the BSpline SH model. Left: The cubic B-spline basis functions, $B_i(G)$, utilized to model the magnitude dependence. The vertical dotted lines indicate the internal knot locations. Unlike piecewise linear functions, cubic B-splines enforce $C^2$ continuity (smoothness in the first and second derivatives) across the magnitude domain. Right: The cubic B-spline basis functions, $B_j(\nu_{\text{eff}})$, used to model the chromatic dependence. }
              \label{fig:bspline_basis}
\end{figure}

\begin{figure}
              \centering
              \subfigure{\includegraphics[width=1.0\linewidth]{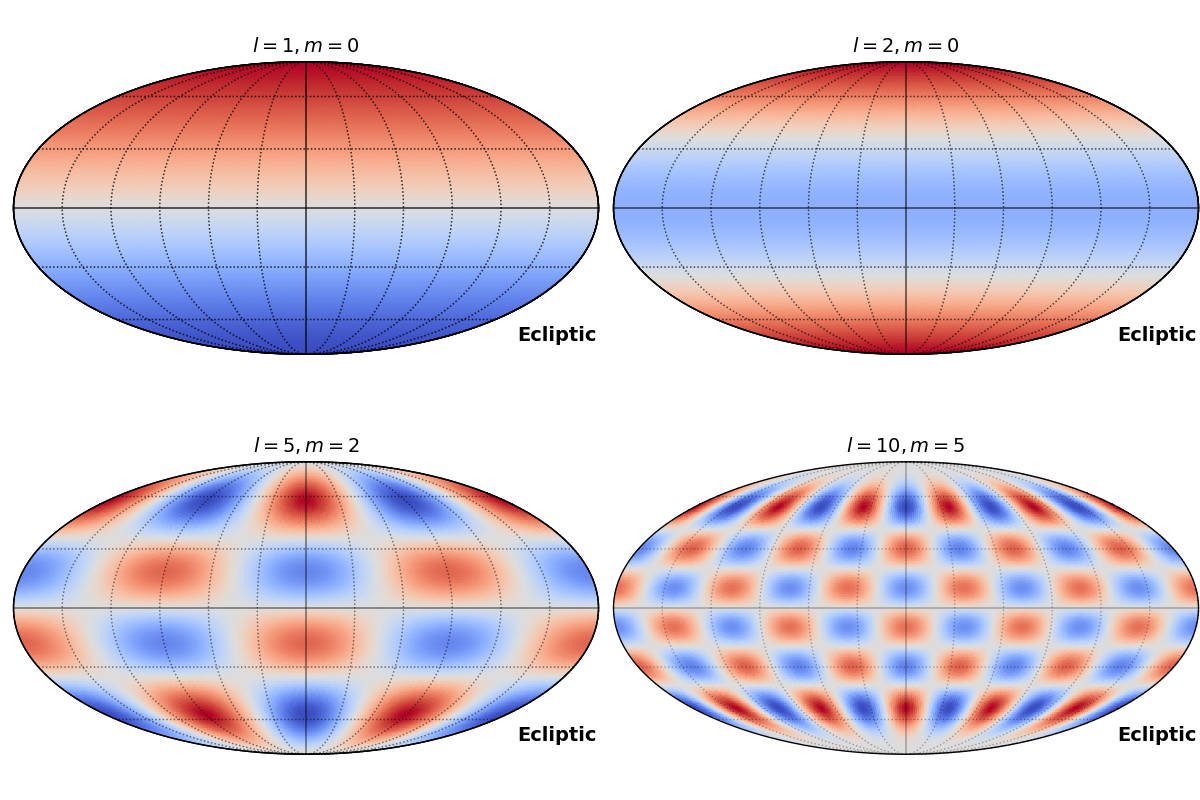}}
              \caption{
              Representative spatial basis functions for the BSpline SH model. The maps display the real component of the Spherical Harmonics, $Y_{lm}(\lambda, \beta)$, in equatorial coordinates (or ecliptic, depending on your implementation), for selected degrees ($l$) and orders ($m$). While low-degree modes (e.g., $l=1, 2$) efficiently capture broad hemispheric asymmetries, even relatively high-degree modes (e.g., $l=10$) exhibit smooth, undulating patterns that are structurally distinct from the sharp, high-frequency striping artifacts characteristic of the Gaia scanning law.}
              \label{fig:sh_basis}
\end{figure}

Figure~\ref{fig:bspline_basis} displays the cubic B-spline bases for magnitude ($G$) and effective wavenumber ($\nu_{\text{eff}}$). The knot placement for the $G$-splines (left panel) is intentionally dense around $G \approx 13$ and $16$, allowing the model to adapt to rapid zero-point variations caused by instrumental transitions (e.g., changes in CCD window classes). To capture spatial variations, the model employs a spherical harmonic expansion up to degree $L_{max} = 10$. Figure~\ref{fig:sh_basis} provides an all-sky visualization of several representative $Y_{lm}$ modes. As discussed in Sect.~\ref{sec:global_limits}, these visualizations highlight the fundamental limitation of this approach: smooth analytical functions on the sphere, even at moderately high degrees, cannot efficiently reproduce the localized, abrupt discontinuities created by the satellite's scanning strategy.

The model coefficients were estimated via a joint least-squares minimization. The loss function $\chi^2_{total}$ incorporates constraints from three distinct datasets:
\begin{equation}
    \chi^2_{total} = \chi^2_{QSO} + W_{bin} \cdot \chi^2_{Bin} + W_{\text{LMC}} \cdot \chi^2_{LMC} + \lambda_{reg} \cdot \|\mathbf{c}\|^2
\end{equation}
where $\chi^2_{QSO}$ represent the squared residuals of quasars, providing an absolute ZP anchor ($\varpi_{true} = 0$). $\chi^2_{Bin}$ represent the squared residuals of the \textit{differential} ZP for wide binaries, constraining the relative shape of the ZP function across different magnitudes and colors. $\chi^2_{\text{eff}}$ represent the squared residuals of LMC member stars. Unlike the L21 approach, we treat LMC stars as having a fixed mean parallax of $\varpi_{\text{LMC}}$ = 20.04 $\mu$as, providing a strong anchor in the bright regime at the South Ecliptic Pole. $W_{bin}$ and $W_{\text{LMC}}$ are weight factors (set to $1.0$ in our final run) used to balance the contribution of the different datasets. $\lambda_{reg} \cdot \|\mathbf{c}\|^2$ is an L2 regularization term (ridge regression) with $\lambda_{reg} = 10^{-4}$ to prevent overfitting and stabilize the solution for high-order terms.

The optimization was performed using the Trust Region Reflective algorithm implementation in \texttt{scipy.optimize.least\_squares}.

\section{Limitations of Global Models}

\label{app:global_limits}

        \begin{figure}
              \centering
              \subfigure{\includegraphics[width=1.0\linewidth]{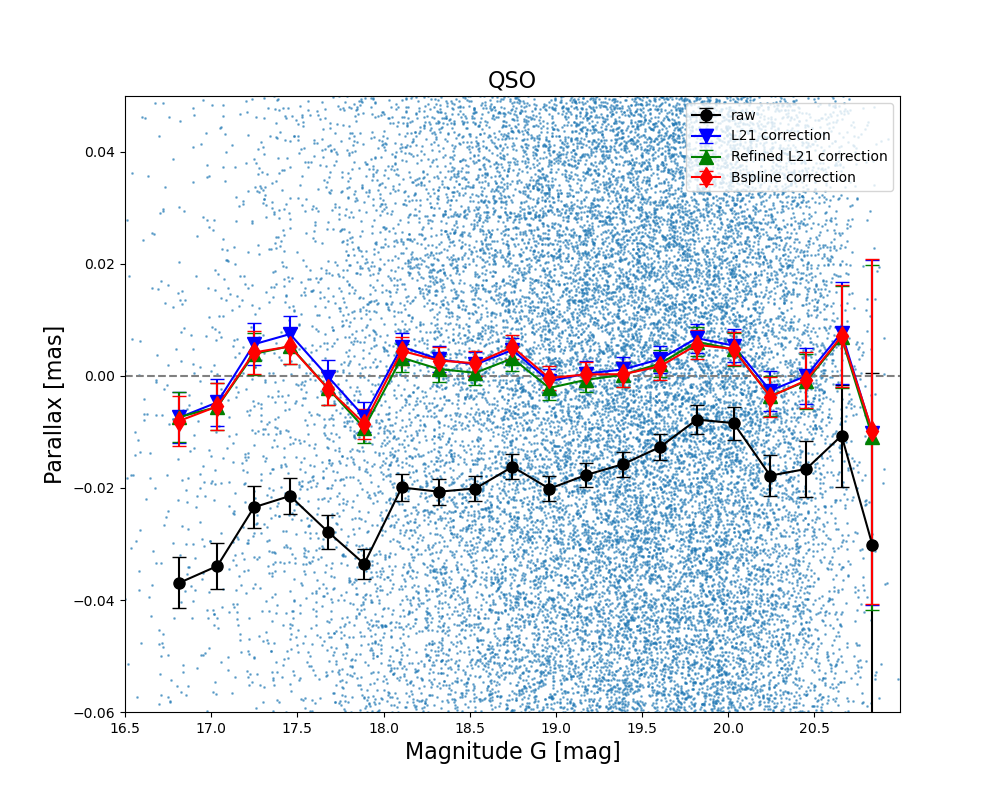}}
              \subfigure{\includegraphics[width=1.0\linewidth]{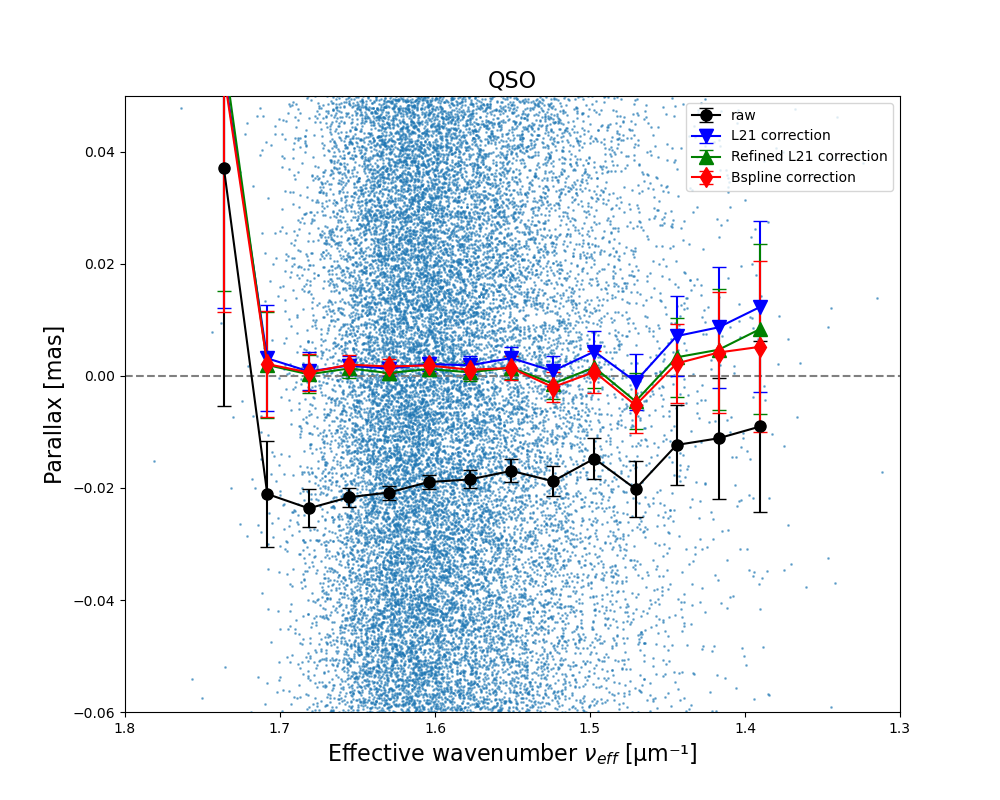}}
              \subfigure{\includegraphics[width=1.0\linewidth]{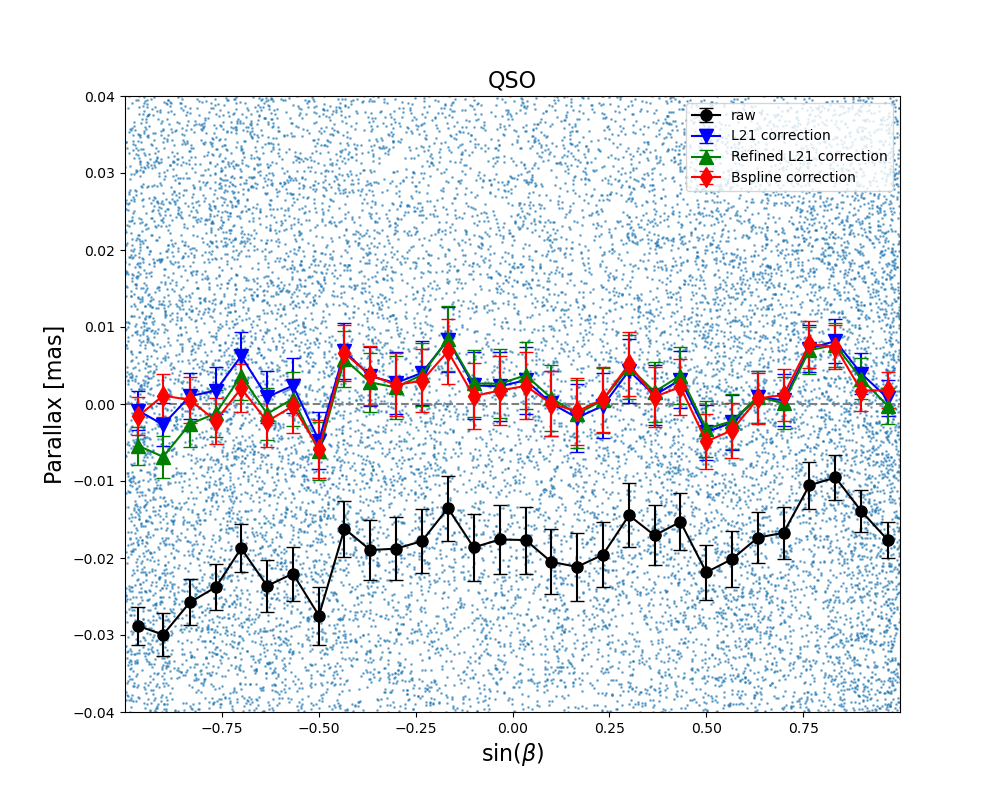}}
              \caption{
               The parallax residuals for the QSOs plotted against $G$ (Top), $\nu_{eff}$ (Middle), and $\sin\beta$ (Bottom), applying three global models: the official L21 model (blue), our Refined L21 model (green), and our experimental BSpline SH model (red). }
              \label{fig:global_comparisons_qso}
        \end{figure}

        \begin{figure}
            \centering
            \subfigure{\includegraphics[width=1.0\linewidth]{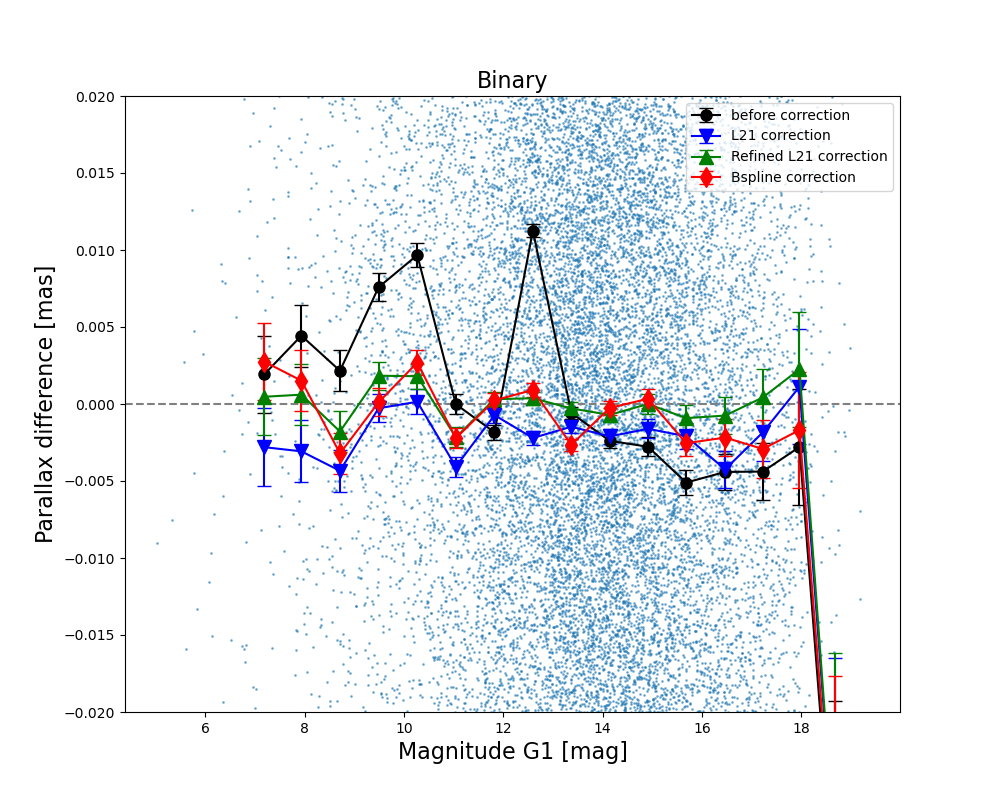}}
             \subfigure{\includegraphics[width=1.0\linewidth]{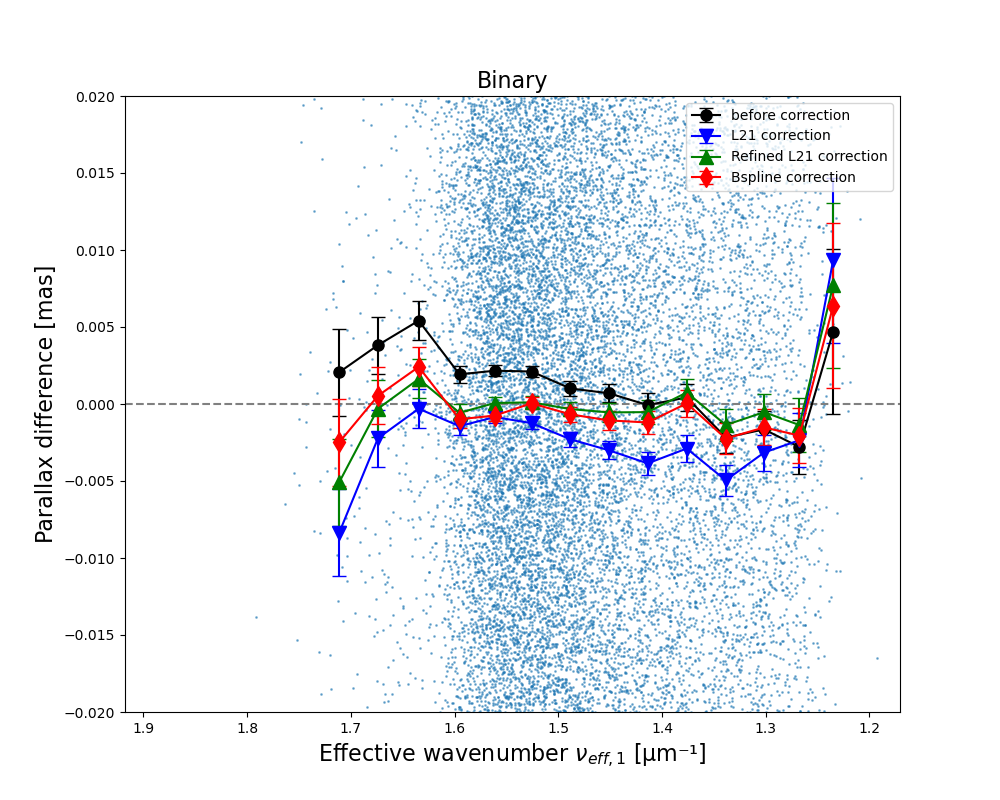}}
             \subfigure{\includegraphics[width=1.0\linewidth]{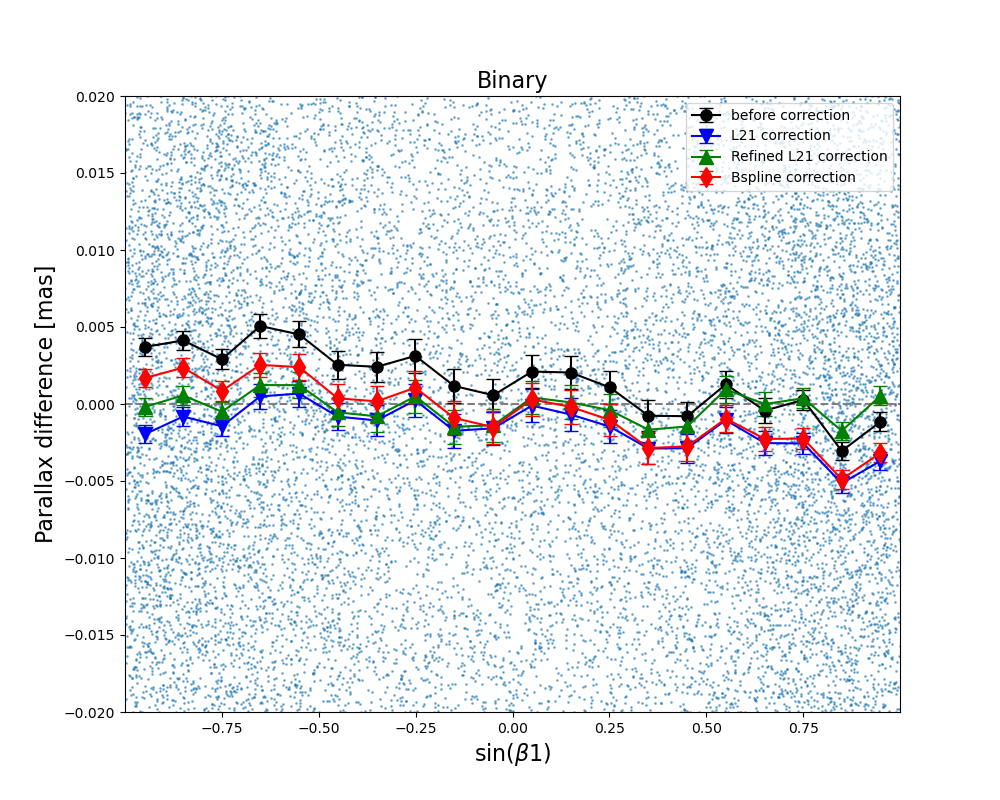}}
            \caption{ 
            The parallax differential residuals for the WBs plotted against $G_1$ (Top), $\nu_{eff1}$ (Middle), and $\sin\beta_1$ (Bottom), applying three global models: the official L21 model (blue), our Refined L21 model (green), and our experimental BSpline SH model (red). }
            \label{fig:global_comparisons_binary}
        \end{figure}
        
        \begin{figure}
            \centering
            \subfigure{\includegraphics[width=1.0\linewidth]{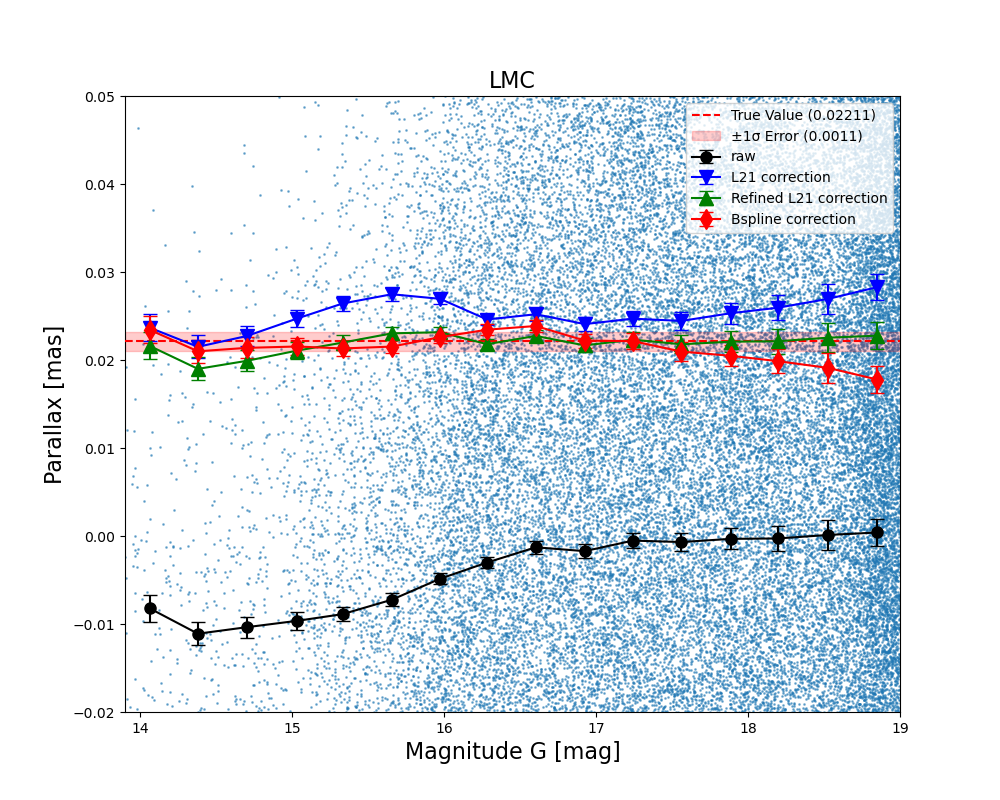}}
            \subfigure{\includegraphics[width=1.0\linewidth]{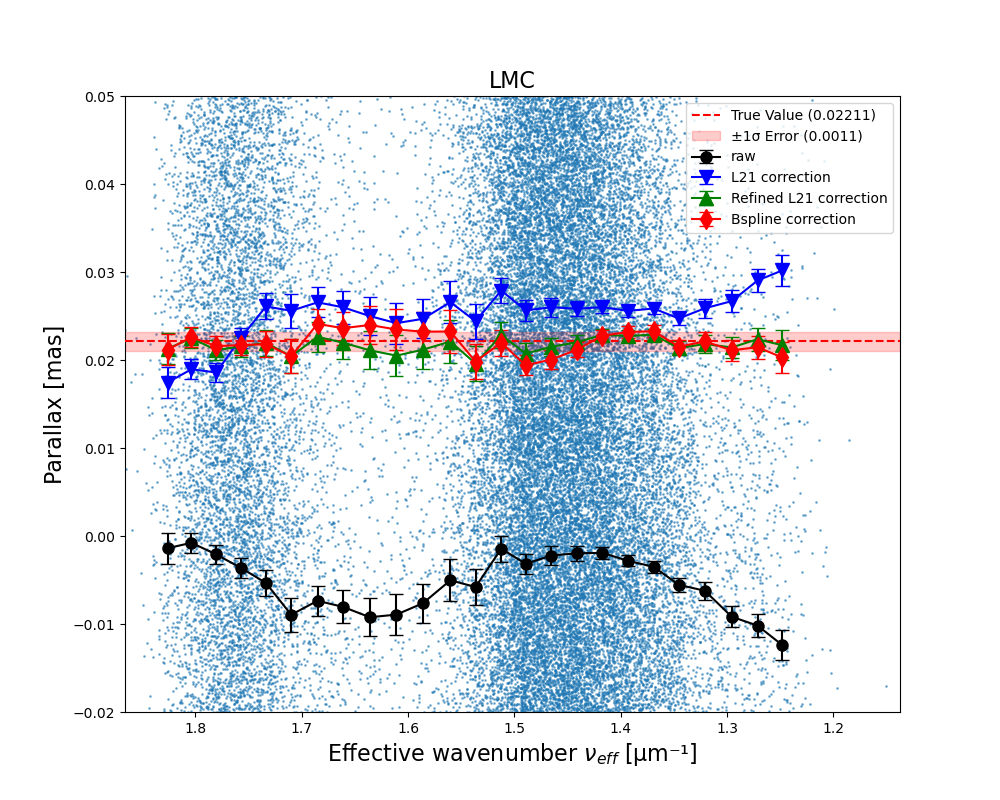}}
            \caption{ The parallax residuals for the LMC sources plotted against $G$ and $\nu_{\text{eff}}$, applying three global models: the official L21 model (blue), Refined L21 model (green), and  BSpline SH model (red). The red shaded area in the LMC plots represents the uncertainty of the LMC mean parallax ($\pm 1.1\,\mu$as)}
            \label{fig:global_comparisons_lmc}
        \end{figure}

    \begin{figure*}
        \centering
        \subfigure{\includegraphics[width=0.45\linewidth]{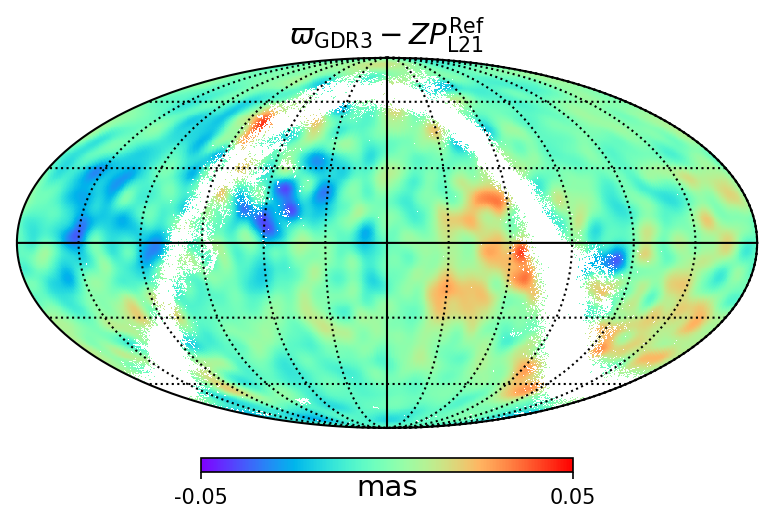}}
        \subfigure{\includegraphics[width=0.45\linewidth]{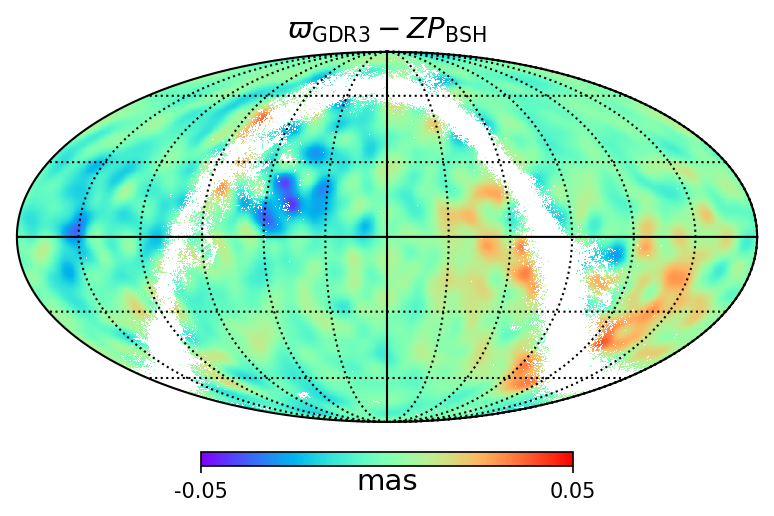}}
        \caption{
        All-sky distribution of parallax residuals for QSOs after applying the Refined L21 (left) and BSH (right) correction. The maps are smoothed with a $8^\circ$ Gaussian kernel. }
        \label{fig:global_comparisons_maps_refinedl21}
    \end{figure*}
    
    \begin{figure*}
        \centering
        
        \subfigure{\includegraphics[width=0.45\linewidth]{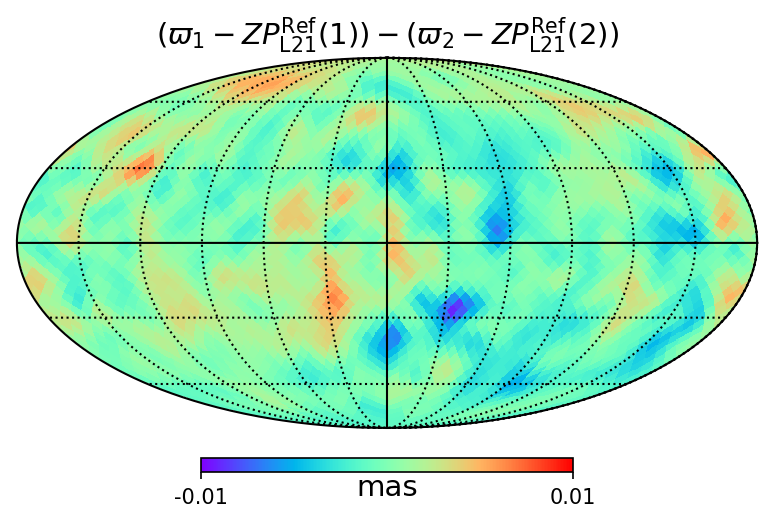}}
        \subfigure{\includegraphics[width=0.45\linewidth]{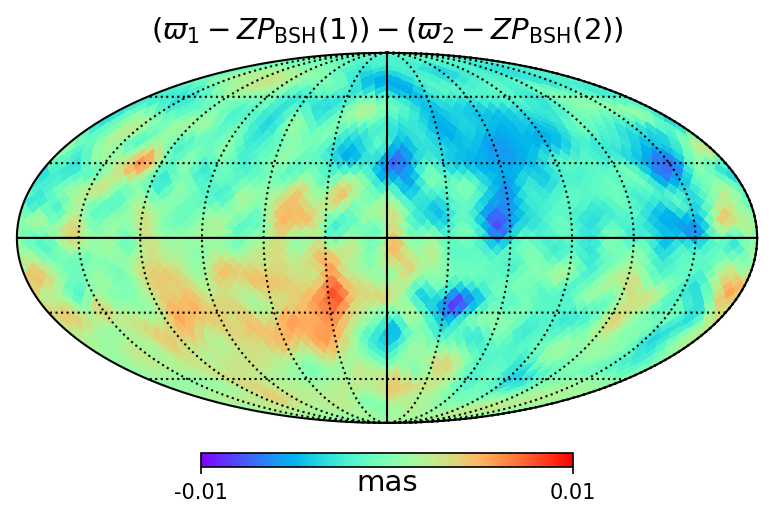}}
        \caption{
        All-sky distribution of differential parallax residuals for binaries after applying Refined L21 (left) and BSH (right) correction. The maps are smoothed with a $10^\circ$ Gaussian kernel. }
        \label{fig:global_comparisons_maps_bsh}
    \end{figure*}

As a first step towards improving the parallax ZP calibration, we investigated two distinct global parametric strategies: (1) refining the coefficients of the standard L21 model using our expanded dataset ("Refined L21"), and (2) constructing a more flexible model based on B-splines and Spherical Harmonics ("BSpline SH"). Both models were trained jointly on quasars, wide binaries, and LMC stars to simultaneously constrain magnitude, color, and spatial dependencies.

Figures~\ref{fig:global_comparisons_qso} and \ref{fig:global_comparisons_binary} compare the performance of these models against the official L21 solution using the QSO and Binary sets.
In terms of parameter dependencies, our Refined L21 model (green triangles) and BSpline SH model (red circles) both achieve a significant improvement in removing the color-dependent trends ($\nu_{\text{eff}}$) seen in the LMC sample, demonstrating the value of including diverse LMC stars in the training.
However, both global models struggle to anchor the absolute zero-point across the full magnitude range. As seen in the LMC magnitude plots (bottom left), they exhibit systematic offsets of $\sim 10-20\,\mu$as for bright stars ($G < 13$), failing to converge to the expected zero level within the intrinsic uncertainty of the LMC distance (grey shaded area).

Crucially, the limitations of global modeling are most evident in the spatial domain. Figures~\ref{fig:global_comparisons_maps_refinedl21} and \ref{fig:global_comparisons_maps_bsh} display the all-sky distribution of residuals after applying corrections from the Refined L21 and BSpline SH models.
Despite the increased flexibility of the BSpline SH model (which includes higher-order spherical harmonics), the residual maps for all three models are remarkably similar. They all exhibit persistent large-scale structures (red/blue patches) and characteristic "striping" patterns associated with the Gaia scanning law, with amplitudes of $\sim 20\,\mu$as.
This convergence of results leads to a robust conclusion: the complicated spatial variations of the zero-point are intrinsic to the scanning law and cannot be effectively captured by smooth global functions, regardless of the specific basis functions chosen. This fundamental limitation necessitates the data-driven, local refinement approach implemented in our Hybrid Strategy.


\bsp	
\label{lastpage}
\end{document}